\documentclass[aps]{revtex4}
\usepackage{eurosym}
\usepackage{amsfonts}
\usepackage{amsmath}
\usepackage{amssymb,epsf}
\usepackage{color}
\usepackage{graphicx}
\usepackage{epstopdf}
\usepackage{float}
\usepackage{caption}
\usepackage{subfig}

\begin{document}

\title{Analytic Electrically Charged Black Holes in $F(R)$-ModMax Theory}
\author{Behzad Eslam Panah$^{1,2,3}$\footnote{%
email address: eslampanah@umz.ac.ir}}
\affiliation{$^{1}$ Department of Theoretical Physics, Faculty of Science, University of
Mazandaran, P. O. Box 47416-95447, Babolsar, Iran}
\affiliation{$^{2}$ ICRANet-Mazandaran, University of Mazandaran, P. O. Box 47416-95447
Babolsar, Iran}
\affiliation{$^{3}$ ICRANet, Piazza della Repubblica 10, I-65122 Pescara, Italy}

\begin{abstract}
Motivated by a new model of nonlinear electrodynamics known as Modified
Maxwell (ModMax) theory, an exact analytical solution for black holes is
obtained by coupling ModMax nonlinear electrodynamics and $F(R)$ gravity.
Then, the effects of the system's parameters ($F(R)$-ModMax gravity
parameters) on the event horizons are analyzed. The obtained black holes
thermodynamic properties in the $F(R)$-ModMax theory are investigated by
extracting their thermodynamic quantities such as Hawking temperature,
electric charge, electric potential, entropy, and also total mass. The first
law of thermodynamics for the system under study is evaluated. Next, by
considering these black holes, the impact of various parameters on both the
local stability and global stability are investigated by examining the heat
capacity and the Helmholtz free energy, respectively. Finally, the
thermodynamic geometry of the black hole in $F(R)$-ModMax gravity is
investigated by applying the thermodynamic metric (the HPEM metric).
\end{abstract}

\maketitle

\section{Introduction}

For several decades, significant interest has been drawn towards $F(R)$ gravity because this modified theory of gravity can describe
the whole universe's evolution by introducing some consistent models \cite%
{NojiriO2003} (see Ref. \cite{NojiriO2011}, for more details). In addition, 
$F(R)$ gravity \cite{F(R)I,F(R)II,F(R)IIb,F(R)III,F(R)IV} can provide
explanations for various phenomena observed in astrophysics and cosmology 
\cite{Mod1,Mod2,Mod3,Mod4,Mod5,Mod6,Mod7,Mod7a,Mod7b,Mod8}. For example,
this theory of gravity can describe the accelerating expansion of our
Universe \cite{Capozziello2002,Carroll2004,Sotiriou2006,Hu2007,Baghram2007},
the existence of the early universe's inflation \cite%
{NojiriO2008,Cognola2008,Elizalde2010}, explain the dark matter \cite%
{NojiriOdark2006,CapozzielloCT2007,CapozzielloH2013}. In addition, $F(R)$
gravity is able to describe the whole sequence of evolution epochs of the
Universe. Additionally, the $F(R)$ gravity theory aligns with both the
Newtonian and post-Newtonian approximations \cite{CapozzielloI,CapozzielloII}%
.

Nonlinear electrodynamics (NED) can describe, for example; i) the virtual
electron-positron pairs' self-interaction \cite{NED1,NED2,NED3}. ii) the NED
field modifies the gravitational redshift around super-strong magnetized
compact objects \cite{NEDMagI,NEDMagII}, iii) removing the singularities due
to both the black hole (which is known as the regular black hole 
\cite{Bardeen1968,Ayon1998,Chinaglia2017,Nojiri2017}) and Big Bang \cite%
{BigBang1,BigBang2,BigBang3}. Notably, the regular black hole is an
object whose spacetime has the horizon, but there is no curvature
singularity. In this regard, regular black holes with multi-horizon
have been studied in general relativity, $F(R)$ gravity,
and Gauss-Bonnet gravity in the presence of NED (see Ref. \cite{Nojiri2017},
for more details).  iv) NED can explain the radiation propagation within
particular substances \cite{NEDI,NEDII,NEDIII,NEDIV}. In addition, the NED
field's effects affect pulsars and higher-magnetized neutron stars \cite%
{NEDP1,NEDP2}. In this regard, Born and Infeld \cite{BI} introduced the
first model of NED in 1934. This particular NED model removes several of the
problems that are encountered in Maxwell's theory, including the elimination
of the electric field's singularity at the center of point particles.
Power-Maxwell NED theory (PM NED) is also an interesting NED theory, whose
Lagrange function is an arbitrary power of Maxwell's Lagrange function \cite%
{PM1,PM2,PM3,PM5}. The PM NED remains invariant when subjected to the
conformal transformations $g_{\mu \nu }\rightarrow \Omega ^{2}g_{\mu \nu }$
and $A_{\mu }\rightarrow A_{\mu }$ (here, $A_{\mu }$ and $g_{\mu \nu }$,
respectively, represent the electrical gauge potential and the metric
tensor). In the framework of PM NED, it was highlighted that point particles
may possess a finite electric field at their center and charges could
potentially have a finite self-energy \cite{RemovePM1,RemovePM2}. The
Maxwell electrodynamics is particularly noteworthy for its duality and scale
invariance. In this regard, in 2020, Bandos, Lechner, Sorokin, and Townsend
introduced a modified Maxwell model (where is known as the ModMax) of
nonlinear duality-invariant conformal electrodynamics. In essence, the
ModMax model of NED exhibits both the duality and also conformal symmetries,
similar to Maxwell's theory \cite{ModMaxI}.

Examining the black holes' exciting features in any gravity can provide
important information about the theory from both theoretical and
observational perspectives. On the other hand, in the gravitational collapse
scenario, all forms of matter, including charged particles, are assimilated
by the black hole. Hence, it becomes crucial to examine the interplay
between the black hole, the linear and nonlinear electromagnetic fields, and
the impact of these fields on the geometry in the context of $F(R)$ gravity.
In this regard, various black holes in the $F(R)$ gravity framework with (or
without) matter field were extracted in Refs. \cite%
{BH1,BH2,BH3,BH4,BH5,BH6,BH7,BH8,BH9,BH10,BH11,BH12,BH13,BH14,BH15,BH16,BH17,BH18,BH19,BH20,BH21,BH22,BH23,BH24,BH25,BH26,BH27,BH28}%
. Furthermore, the field equations of $F(R)$ gravity pose a challenge as
they are intricate 4th order differential equations, making it arduous to
derive analytical solutions for black holes, especially in the presence of
matter field such as NED theories. Based on the specified characteristics of 
$F(R)$ gravity and the ModMax-NED theory (which includes the same symmetries
as Maxwell's theory, i.e. electromagnetic duality and conformal invariance),
we are interested in extracting analytical solutions for black holes in this
theory.

By connecting the geometrical quantities such as surface gravity to
temperature and horizon area to entropy, the black hole as a thermodynamic
system is studied by Bekenstein and Hawking \cite{Bekenstein1973,Hawking1974}%
. Next, for non-rotating and uncharged black holes, $dM=TdS$ was established
as the first law of black hole thermodynamics \cite{Hawking1974,Bardeen1973}%
. In this regard, studying the thermodynamical properties of black holes,
especially phase transition has assimilated much interest \cite%
{The1,The2,The3,The4,The5,The6,The7,The8,The9,The10,The11,The12,The13,The14,The15,The16,The17,The18,The19,The20,The21,The22,The23,The24}%
.

There exist multiple approaches to peruse the phase transition of black
holes, such as examining their heat capacity and utilizing geometrical
thermodynamics (GTD). The study of the heat capacity reveals two distinctive
points which are known as bound and phase transition points. i) the bound
point determines when the numerator of heat capacity (or temperature) is
zero. At this point, the sign of the temperature changes, and we can
separate the non-physical and physical black holes. ii) the heat capacity
divergences determine the phase transition points. Furthermore, the heat
capacity's positivity (negativity) ensures the system's thermal
(in)stability. So, the study of the heat capacity gives us information about
the thermal stability, the phase transition, and physical bound points of
the system \cite{HeatC1,HeatC2,HeatC3}.

GTD is an alternative approach used to evaluate the phase transition of
black holes. GTD involves constructing a thermodynamic metric using
thermodynamic potentials, such as entropy or internal energy, and their
derivatives with respect to the system's extensive parameters. The
divergences of the Ricci scalar of this thermodynamic metric provide
insights into the phase transition points. Various thermodynamic metrics,
including those proposed by Weinhold \cite{WeinholdI,WeinholdII}, Ruppeiner 
\cite{RuppeinerI,RuppeinerII,RuppeinerIII}, Quevedo \cite{QuevedoI,QuevedoII}%
, and Hendi-Panahiyan-EslamPanah-Momennia (referred to as HPEM's metric) 
\cite{HPEM,HPEM1,HPEM2,HPEM3}, have been introduced to investigate this
phenomenon. However, it has been observed the Ricci scalar's singularities
in the Weinhold and Ruppeiner metrics do not align with the singularities of
heat capacity, making them unsuitable for explaining the thermodynamic
properties of different black holes. This issue arises from the lack of
Legendre invariance in these metrics. To address this problem, Quevedo
introduced a new type of thermodynamic metric, known as Quevedo's metric,
which is invariant under Legendre transformations\cite{QuevedoI,QuevedoII}.
Other thermodynamic metrics have also been proposed in the literature \cite%
{GTD1,GTD2,GTD3,GTD4}, but they have their own limitations. In this context,
Hendi-Panahiyan-EslamPanah-Momennia developed a new metric, HPEM's metric,
which overcomes the shortcomings of previous thermodynamic metrics and
effectively distinguishes between phase transition and bound points \cite%
{HPEM,HPEM1,HPEM2,HPEM3}. Therefore, we consider HPEM's metric to study the
bound and phase transitions points, and also stability conditions for the
obtained black holes in $F(R)-$ModMax theory.

This paper follows the following outline. The subsequent section presents an
introduction to the field equations in $F(R)$ gravity in the presence of
ModMax NED. The electric black hole solution in $F(R)$-ModMax gravity will
be derived and the influence of the parameters on these black holes will be
evaluated in Section III. Section IV will cover the thermodynamic quantities
and an examination of the first law of thermodynamics. The forthcoming
section will investigate the impact of various parameters on both local and
global stability through the utilization of heat capacity and Helmholtz free
energy. This analysis will be carried out in Section V. Section VI will
focus on the study of the phase transition and the physical limitation
points for the extracted black holes within the framework of GTD using
HPEM's metric. The final section will be dedicated to concluding remarks.

\section{Field equations in F(R)-ModMax theory}

In this study, we investigate the coupling of the ModMax field (as the
source of matter) with $F(R)$ gravity. In four-dimensional spacetime, the
action of $F(R)$-ModMax theory is given by 
\begin{equation}
\mathcal{I}_{F(R)}=\frac{1}{16\pi }\int_{\partial \mathcal{M}}d^{4}x\sqrt{-g}%
\left[ F(R)-4\mathcal{L}\right] ,  \label{actionF(R)}
\end{equation}%
where $F(R)=R+f\left( R\right) $, in which $R$ and $f\left( R\right) $,
respectively, are scalar curvature and a function of scalar curvature. In
this paper, we consider the Newtonian gravitational constant and the speed
of light equal to $1$, i.e., $G=c=1$. The second term in the above action is
devoted to the ModMax Lagrangian ($\mathcal{L}$), which is defined \cite%
{ModMaxI,ModMaxII} 
\begin{equation}
\mathcal{L}=\frac{1}{2}\left( \mathcal{S}\cosh \gamma -\sqrt{\mathcal{S}^{2}+%
\mathcal{P}^{2}}\sinh \gamma \right) ,  \label{ModMaxL}
\end{equation}%
where $\gamma $ is known as the parameter of ModMax theory. $\gamma $ is a
dimensionless parameter. Also, $\mathcal{S}$, and $\mathcal{P}$ are,
respectively, a true scalar, and a pseudoscalar, which are defined in the
following forms 
\begin{equation}
\mathcal{S}=\frac{\mathcal{F}}{2},~~~\&~~~\mathcal{P}=\frac{\widetilde{%
\mathcal{F}}}{2},
\end{equation}%
and $\mathcal{F}=F_{\mu \nu }F^{\mu \nu }$ is the Maxwell invariant ($F_{\mu
\nu }=\partial _{\mu }A_{\nu }-\partial _{\nu }A_{\mu }$ (where $A_{\mu }$
is the gauge potential) is the electromagnetic tensor). In addition, $%
\widetilde{\mathcal{F}}$ equals to $F_{\mu \nu }\widetilde{F}^{\mu \nu }$,
where $\widetilde{F}^{\mu \nu }=\frac{1}{2}\epsilon _{\mu \nu }^{~~~\rho
\lambda }F_{\rho \lambda }$. This nonlinear electromagnetic theory reduces
to Maxwell's theory, when $\gamma =0$. Moreover, $g=det(g_{\mu \nu })$ is
the determinant of metric tensor $g_{\mu \nu }$, in the action (\ref%
{actionF(R)}).

In this work, we are interested in considering the electrically charged case
of the ModMax theory. In other words, we want to obtain electrical charged
black hole solutions by coupling $F(R)$ theory and the ModMax nonlinear
electrodynamics theory. Therefore, we have to consider $\mathcal{P}=0$ in
the above equations. For this purpose, we are able to extract the equations
of motion of $F(R)$-ModMax theory of gravity, which lead to 
\begin{eqnarray}
R_{\mu \nu }\left( 1+f_{R}\right) -\frac{g_{\mu \nu }F(R)}{2}+\left( g_{\mu
\nu }\nabla ^{2}-\nabla _{\mu }\nabla _{\nu }\right) f_{R} &=&8\pi \mathrm{T}%
_{\mu \nu },  \label{EqF(R)1} \\
&&  \notag \\
\partial _{\mu }\left( \sqrt{-g}\widetilde{E}^{\mu \nu }\right) &=&0,
\label{EqF(R)2}
\end{eqnarray}%
where $f_{R}=\frac{df(R)}{dR}$. Also, $\mathrm{T}_{\mu \nu }$ defines as the
energy-momentum tensor, which is given by 
\begin{equation}
4\pi \mathrm{T}^{\mu \nu }=\left( F^{\mu \sigma }F_{~~\sigma }^{\nu
}e^{-\gamma }\right) -e^{-\gamma }\mathcal{S}g^{\mu \nu },  \label{eq3}
\end{equation}%
and $\widetilde{E}_{\mu \nu }$ in Eq. (\ref{EqF(R)2}), is defined as 
\begin{equation}
\widetilde{E}_{\mu \nu }=\frac{\partial \mathcal{L}}{\partial F^{\mu \nu }}%
=2\left( \mathcal{L}_{\mathcal{S}}F_{\mu \nu }\right) ,  \label{eq3b}
\end{equation}%
where $\mathcal{L}_{\mathcal{S}}=\frac{\partial \mathcal{L}}{\partial 
\mathcal{S}}$. So, the ModMax field equation (Eq. (\ref{EqF(R)2})) for the
electrically charged case reduces to 
\begin{equation}
\partial _{\mu }\left( \sqrt{-g}e^{-\gamma }F^{\mu \nu }\right) =0.
\label{Maxwell Equation}
\end{equation}

\section{Black hole solutions in $F(R)-$ModMax theory}

We consider a static spherically symmetric spacetime as 
\begin{equation}
ds^{2}=-g(r)dt^{2}+\frac{dr^{2}}{g(r)}+r^{2}\left( d\theta ^{2}+\sin
^{2}\theta d\varphi ^{2}\right) ,  \label{Metric}
\end{equation}%
in which $g(r)$ defines as the metric function that we must find.

In general, the equations governing $F(R)$ gravity with a nonlinear matter
field (Eq. (\ref{EqF(R)1})) are intricate. Therefore, deriving a precise
analytical solution is a challenging task. To resolve this problem, one can
consider the traceless energy-momentum tensor for the nonlinear matter field
(like the ModMax field), one can extract an exact analytical solution from $%
F(R)$ gravity coupled to a nonlinear matter field. So, to get the solution
for a black hole with constant curvature in $F(R)$ theory of gravity coupled
to the ModMax field, it is necessary for the trace of the stress-energy
tensor $\mathrm{T}_{\mu \nu }$ to be zero \cite{Rcont1,Rcont2}. Assuming the
constant scalar curvature $R=R_{0}=$ constant, then the trace of the
equation (\ref{EqF(R)1}) turns to 
\begin{equation}
R_{0}\left( 1+f_{R_{0}}\right) -2\left( R_{0}+f(R_{0})\right) =0,
\label{R00}
\end{equation}%
where $f_{R_{0}}=$ $f_{R_{\left\vert _{R=R_{0}}\right. }}$. We can solve the
equation (\ref{R00}) to obtain $R_{0}$ which leads to 
\begin{equation}
R_{0}=\frac{2f(R_{0})}{f_{R_{0}}-1}.  \label{R0}
\end{equation}

By replacing Eq. (\ref{R0}) within Eq. (\ref{EqF(R)1}), the $F(R)$-ModMax
theory's equations of motion can be found in the following format 
\begin{equation}
R_{\mu \nu }\left( 1+f_{R_{0}}\right) -\frac{g_{\mu \nu }}{4}R_{0}\left(
1+f_{R_{0}}\right) =8\pi \mathrm{T}_{\mu \nu }.  \label{F(R)Trace}
\end{equation}

It is notable that, the equation of motion in $F(R)$-ModMax theory (\ref%
{F(R)Trace}) reduces to GR-ModMad theory of graviry when $f_{R_{0}}=0$.

To obtain a radial electric field, we take the following form for the gauge
potential 
\begin{equation}
A_{\mu }=h\left( r\right) \delta _{\mu }^{t},
\end{equation}%
By utilizing the provided gauge potential and equations (\ref{Maxwell
Equation}) and (\ref{Metric}), we are able to derive the subsequent
differential equation. 
\begin{equation}
rh^{\prime \prime }(r)+2h^{\prime }(r)=0,  \label{hh}
\end{equation}%
where, respectively, the prime and double prime represent the first and
second derivatives of $r$. The solution of the equation (\ref{hh}) yields 
\begin{equation}
h(r)=-\frac{q}{r},  \label{h(r)}
\end{equation}%
where $q$ represents an integration constant that is associated with the
electric charge.

We are now able to obtain precise analytical solutions for the metric
function $g\left( r\right) $ by taking into account the metric (\ref{Metric}%
), the derived $h(r)$, and the field equations (\ref{F(R)Trace}).
Consequently, we derive the subsequent set of differential equations 
\begin{eqnarray}
eq_{tt} &=&eq_{rr}=rg^{\prime \prime }(r)+2g^{\prime }(r)+\frac{rR_{0}}{2}-%
\frac{2q^{2}e^{-\gamma }}{r^{3}\left( 1+f_{R_{0}}\right) },  \label{eq11} \\
&&  \notag \\
eq_{\theta \theta } &=&eq_{\varphi \varphi }=rg^{\prime }(r)+g\left(
r\right) -1+\frac{r^{2}R_{0}}{4}+\frac{q^{2}e^{-\gamma }}{r^{2}\left(
1+f_{R_{0}}\right) },  \label{eq22}
\end{eqnarray}%
in which $eq_{tt}$, $eq_{rr}$, $eq_{\theta \theta }$ and $eq_{\varphi
\varphi }$, respectively, are the components of $tt$, $rr$, $\theta \theta $
and $\varphi \varphi $ of field equations (\ref{F(R)Trace}). By utilizing
the aforementioned differential equations, we can obtain a precise solution
for the constant scalar curvature ($R=R_{0}$= constant). After careful
consideration and performing several calculations, the metric function can
be expressed in the subsequent form 
\begin{equation}
g(r)=1-\frac{m_{0}}{r}-\frac{R_{0}r^{2}}{12}+\frac{q^{2}e^{-\gamma }}{\left(
1+f_{R_{0}}\right) r^{2}},  \label{g(r)F(R)}
\end{equation}%
where $m_{0}$ is an integration constant. It is noteworthy that this
constant of integration is connected to the black hole's geometric mass.
Furthermore, any of the field equations (\ref{F(R)Trace}) is satisfied by
the obtained solution (\ref{g(r)F(R)}). We should limit ourselves to $%
f_{R_{0}}\neq -1$ in order to have physical solutions. In addition, we can
see the effect of ModMax's theory in the fourth term of the solution (\ref%
{g(r)F(R)}), and the effect of $F(R)$ gravity both in the third and fourth
terms of the metric function (\ref{g(r)F(R)}). Notably, Reissner-Nordstr\"{o}%
m-(A)dS black hole is covered by considering $f_{R_{0}}=0$, $R_{0}=4\Lambda $
and $\gamma =0$, i.e., 
\begin{equation}
g(r)=1-\frac{m_{0}}{r}-\frac{\Lambda r^{2}}{3}+\frac{q^{2}}{r^{2}}.
\end{equation}

Here, we study the Kretschmann scalar ($R_{\alpha \beta \gamma \delta
}R^{\alpha \beta \gamma \delta }$) in order to find the singularity of
spacetime. Indeed, this quantity gives us information about the existence of
the singularity in spacetime. For this purpose, we calculate the Kretschmann
scalar of the spacetime (\ref{Metric}), which is 
\begin{equation}
R_{\alpha \beta \gamma \delta }R^{\alpha \beta \gamma \delta }=g^{\prime
\prime ^{2}}(r)+\frac{4g^{\prime ^{2}}(r)}{r^{2}}+\frac{4}{r^{4}}-\frac{%
8g\left( r\right) }{r^{4}}+\frac{4g^{2}(r)}{r^{4}},  \label{K}
\end{equation}%
and by replacing the metric function (\ref{g(r)F(R)}) within Eq. (\ref{K}),
we have 
\begin{equation}
R_{\alpha \beta \gamma \delta }R^{\alpha \beta \gamma \delta }=\frac{%
R_{0}^{2}}{6}+\frac{12m_{0}^{2}}{r^{6}}-\frac{48m_{0}q^{2}e^{-\gamma }}{%
\left( 1+f_{R_{0}}\right) r^{7}}+\frac{56q^{4}e^{-2\gamma }}{\left(
1+f_{R_{0}}\right) ^{2}r^{8}}.
\end{equation}%
The obtained Kretschmann scalar includes three important points, which are

i) It diverges at $r=0$, i.e., 
\begin{equation}
\lim_{r\longrightarrow 0}R_{\alpha \beta \gamma \delta }R^{\alpha \beta
\gamma \delta }\longrightarrow \infty ,
\end{equation}
there exists a curvature singularity situated at the coordinate $r=0$.

ii) The Kretschmann scalar is finite for $r\neq 0$.

iii) The effect of ModMax's parameter appears in the Kretschmann scalar.
Although, the divergence of the electrical field is removed by considering $%
\gamma \longrightarrow \infty $, this limit cannot remove the curvature
singularity at $r=0$. In other words, in the limit $\gamma \longrightarrow
\infty $, we have $\lim_{r\rightarrow 0}R_{\alpha \beta \gamma
\delta}R^{\alpha \beta \gamma \delta }\longrightarrow \infty $.

The asymptotical behavior of the Kretschmann scalar and the metric function
are given by 
\begin{eqnarray}
\lim_{r\longrightarrow \infty }R_{\alpha \beta \gamma \delta }R^{\alpha
\beta \gamma \delta } &\longrightarrow &\frac{R_{0}^{2}}{6},  \notag \\
&& \\
\lim_{r\longrightarrow \infty }g\left( r\right) &\longrightarrow &-\frac{%
R_{0}r^{2}}{12}  \notag
\end{eqnarray}%
where indicate that the spacetime will be asymptotically (A)dS, when we
consider $R_{0}=4\Lambda $, and $\Lambda >0$ ($\Lambda <0$). It is notable
that, the asymptotical behavior is independent of $\gamma $. In other words,
the parameter of ModMax does not affect the asymptotical behavior of the
spacetime.

In this context, our objective is to determine the real roots of the
acquired metric function (\ref{g(r)F(R)}) because these roots can give us
information about the horizons (inner and outer horizons) of the solution.
The objects known as black holes possess a curvature singularity situated at 
$r=0$, which is concealed by a minimum of one event horizon. Notably, we can
have black holes without the event horizon, which is known as the naked
singularities.

\begin{figure}[tbph]
\centering
\includegraphics[width=0.35\linewidth]{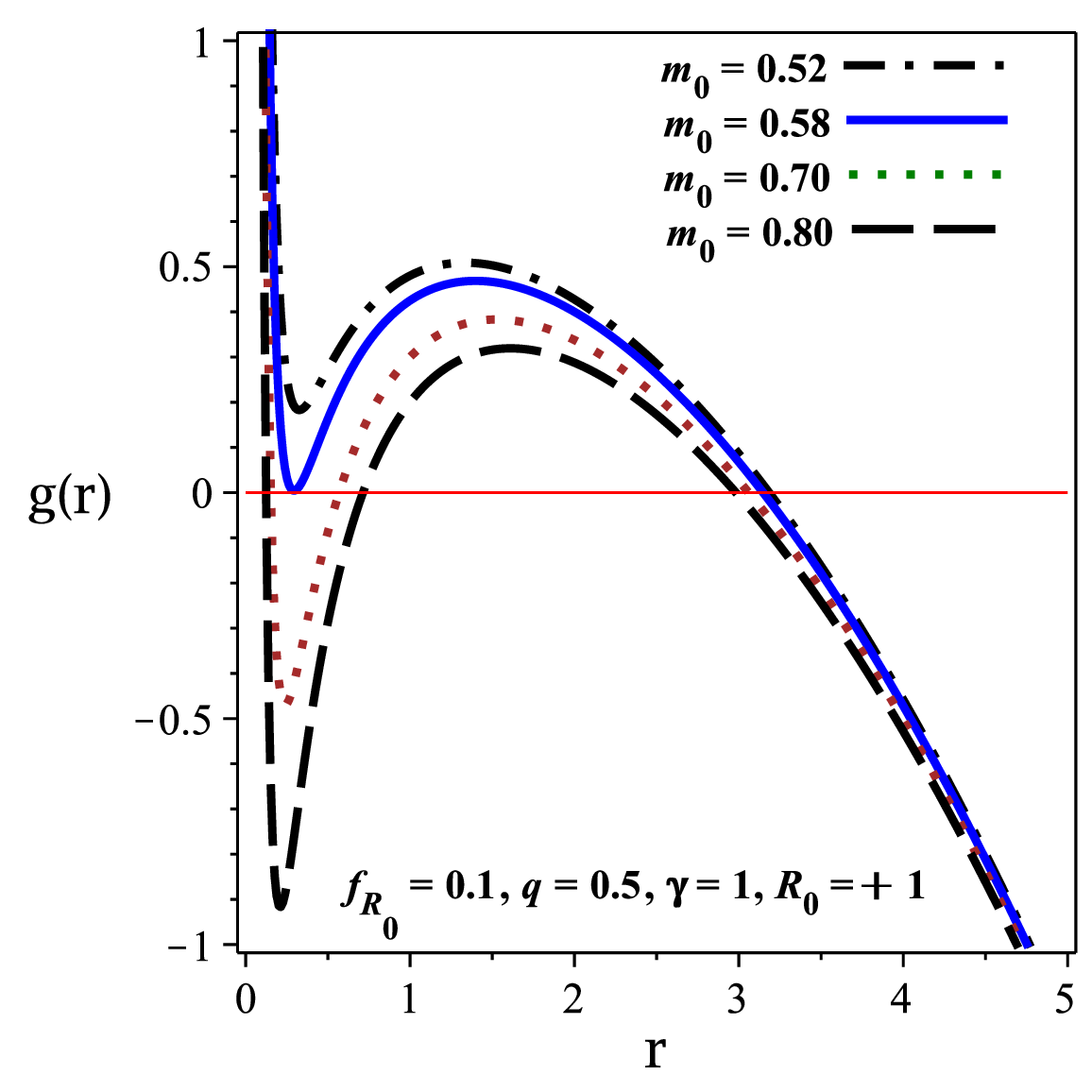} \includegraphics[width=0.35%
\linewidth]{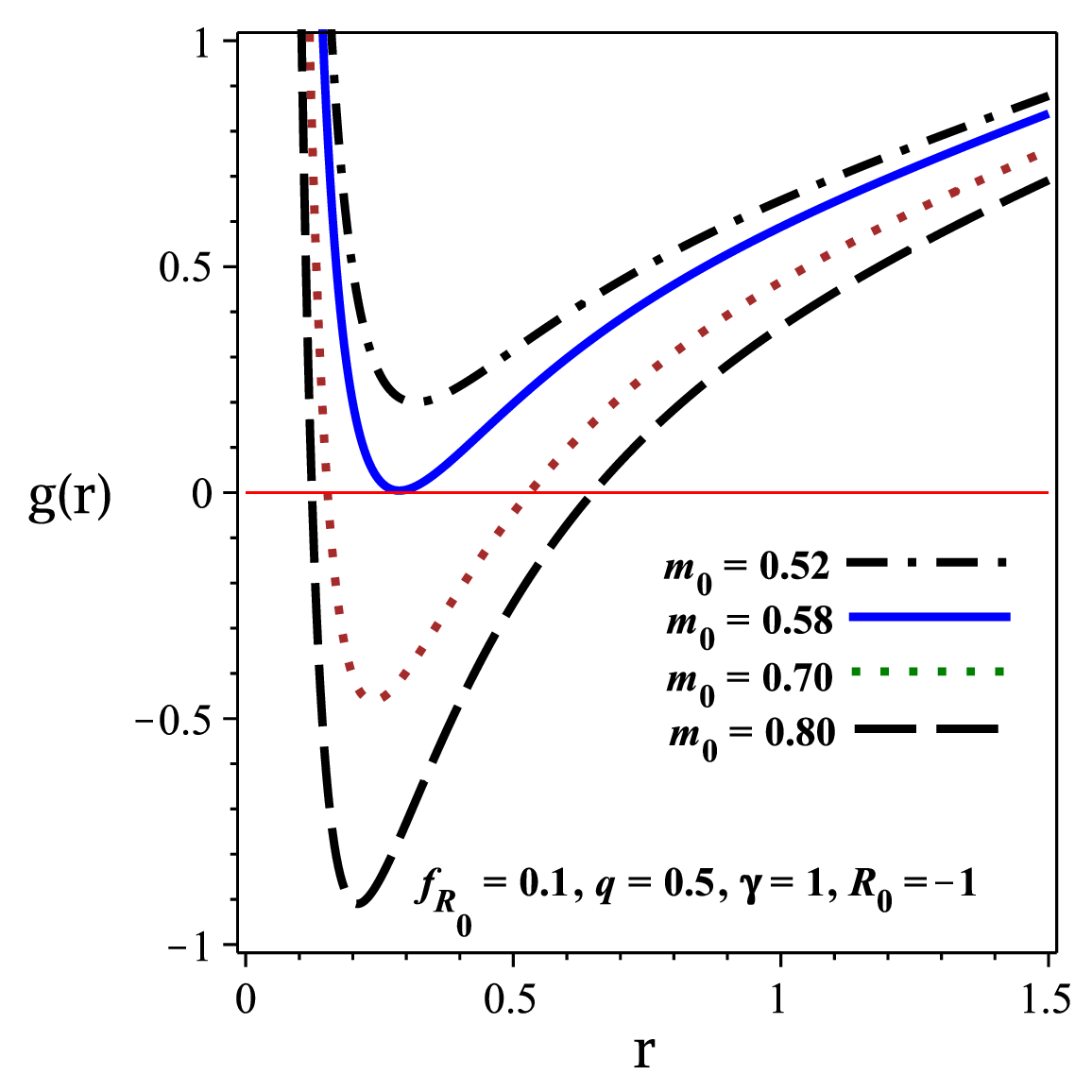}\newline
\caption{The function $g(r)$ is plotted against $r$ for various parameter
values, with the left panels corresponding to $R_{0}=1$ and the right panels
corresponding to $R_{0}=-1$.}
\label{Fig1}
\end{figure}

To find the roots, it is better to solve the metric function. Here, the
metric function is a fourth-order function of $r$, and it is not easy to get
an exact solution. Therefore, the metric function is plotted against the
variable $r$ in Fig. \ref{Fig1} to obtain these roots. As shown in Fig. \ref%
{Fig1}, for $R_{0}>0$ (or dS case if we consider $R_{0}=4\Lambda $), we
encounter three different cases. In the first case, there are three roots,
which are an inner root, an event horizon, and an outer root (cosmological
horizon), respectively. In the second case, there are two roots (extreme
case and cosmological horizon). In the third case, one root (cosmological
horizon) exists. For $R_{0}<0$ (or AdS case if we consider $R_{0}=4\Lambda $%
), the solution may have three different cases which are: i) two roots (an
inner horizon and an event horizon). ii) one root (extreme case). iii) naked
singularity. By modifying certain parameters, it is possible to achieve an
event horizon that encompasses the singularity located at $r=0$. The results
validate that the solution acquired in Eq. (\ref{g(r)F(R)}) may be
associated with the black hole solution in $F(R)$-ModMax theory.

\begin{figure}[tbph]
\centering
\includegraphics[width=0.35\linewidth]{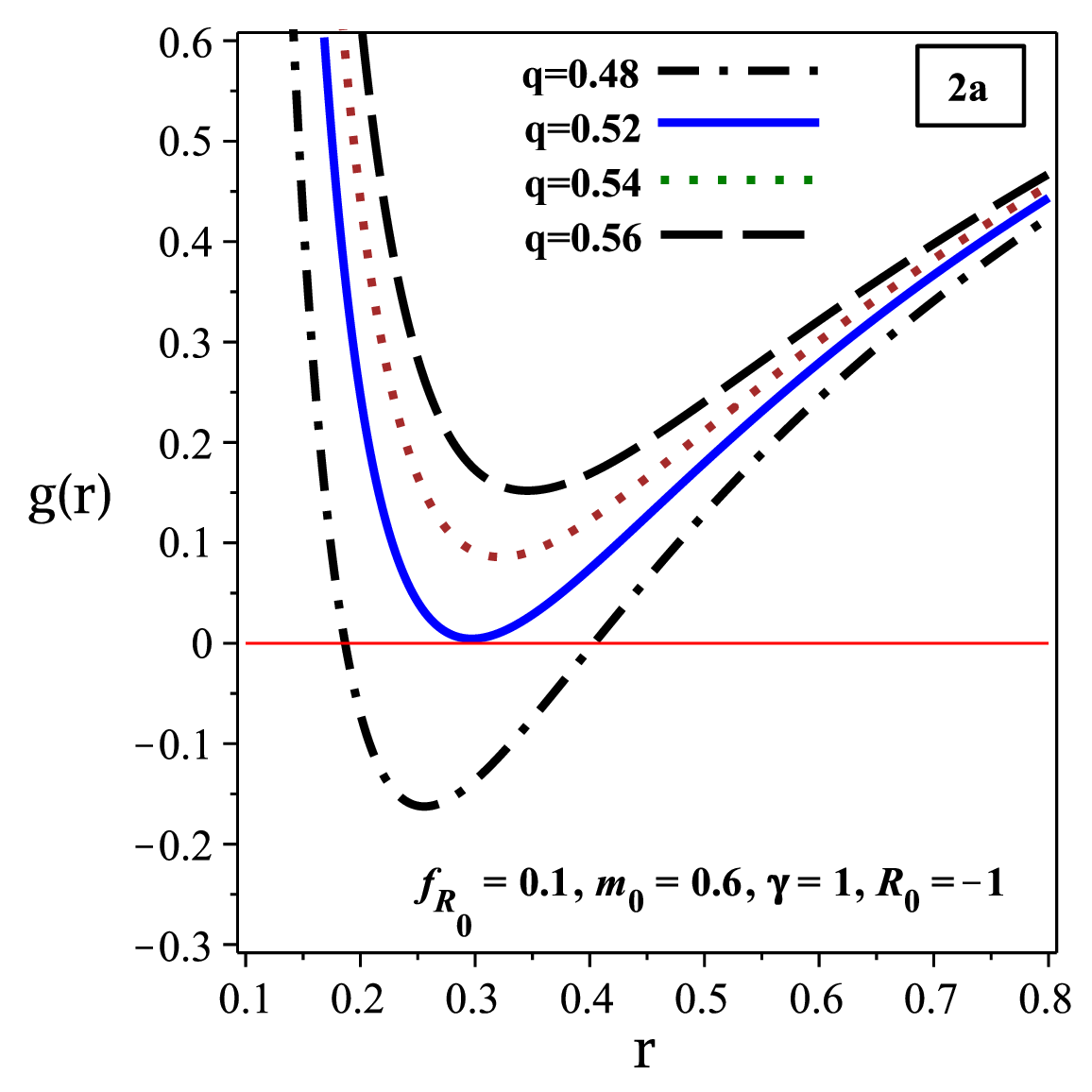} \includegraphics[width=0.35%
\linewidth]{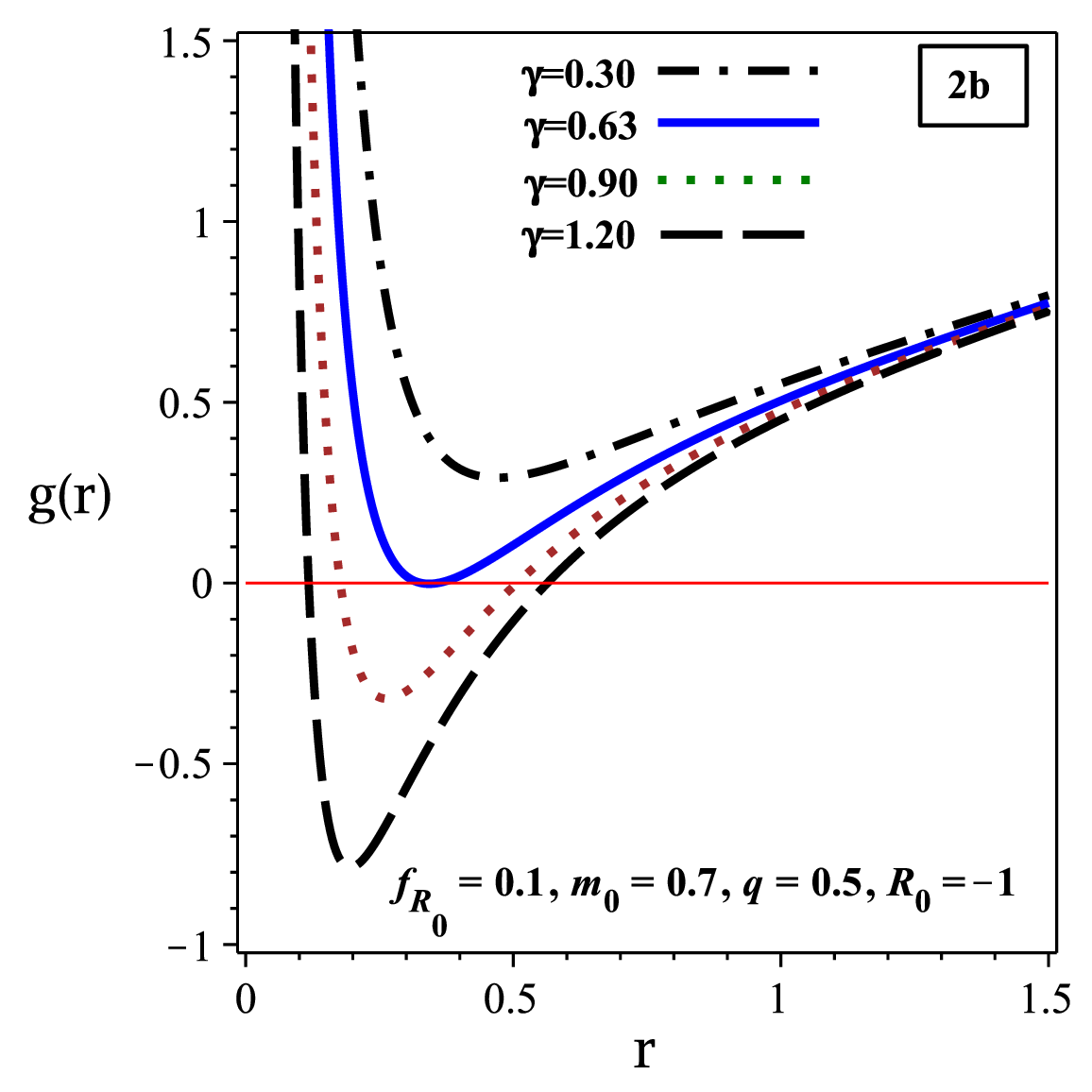}\newline
\includegraphics[width=0.35\linewidth]{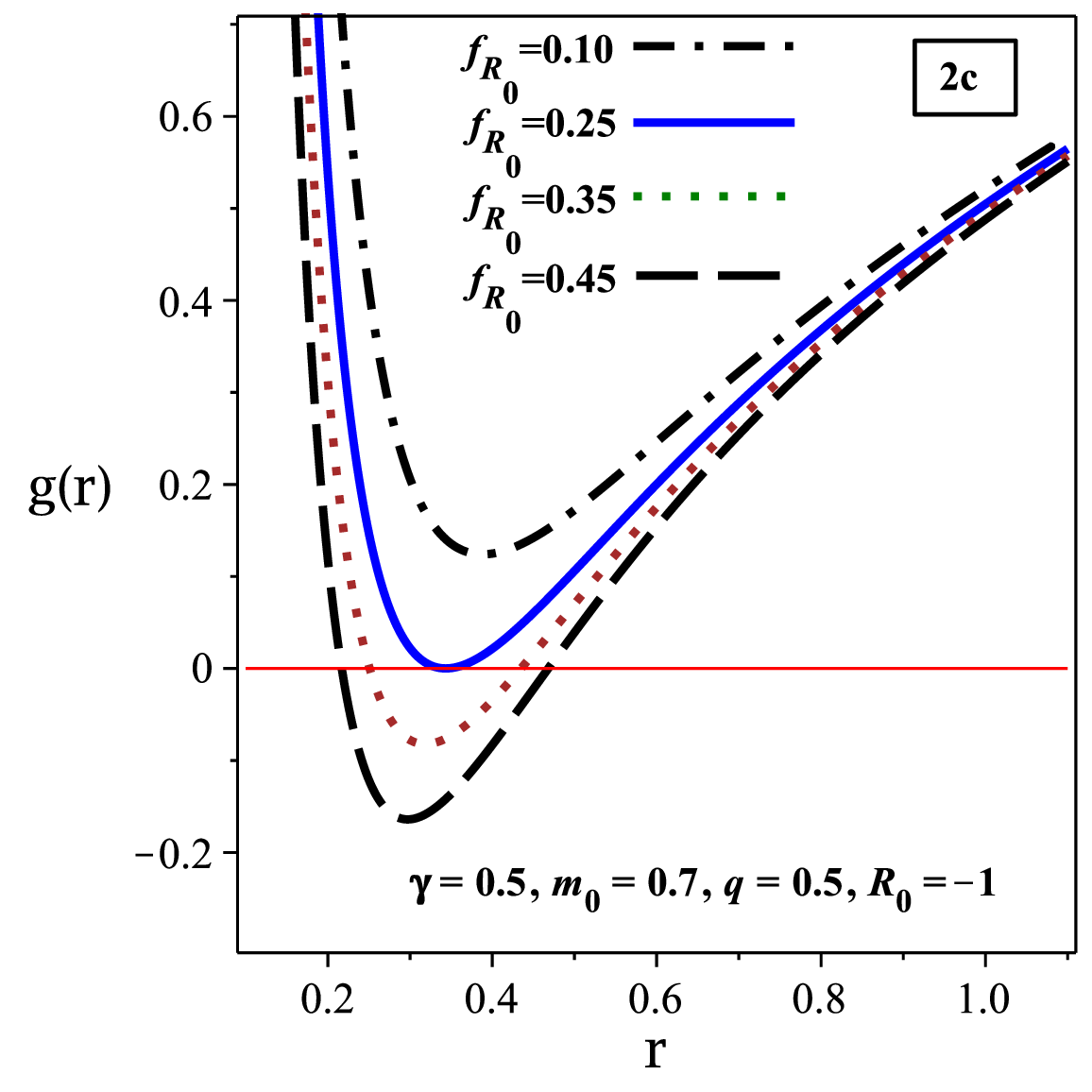} \includegraphics[width=0.35%
\linewidth]{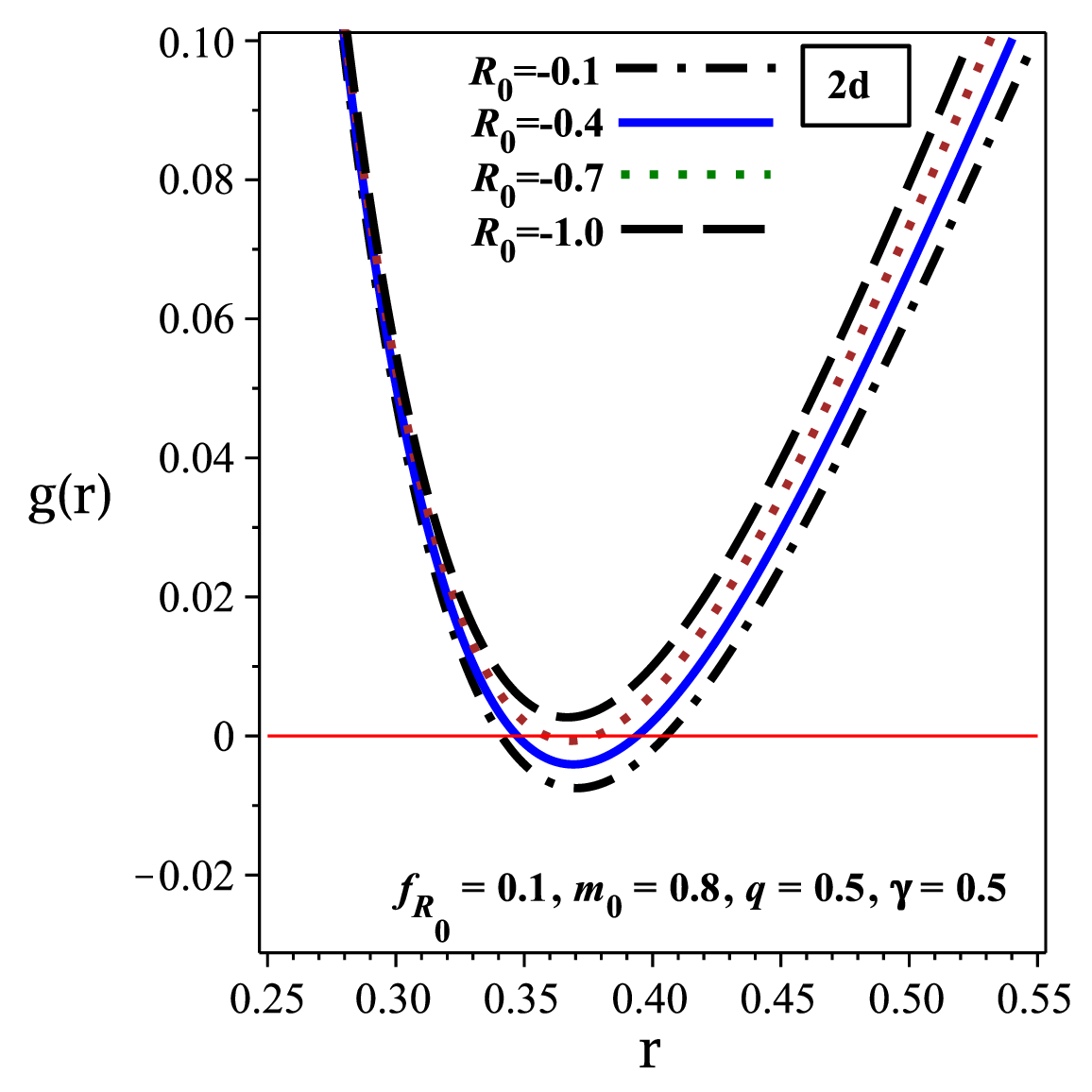}\newline
\caption{The function $g(r)$ versus $r$ is plotted for various parameter
values.}
\label{Fig2}
\end{figure}

Now, we have the opportunity to examine the impact of the parameters in $F(R)
$-ModMax theory on the event horizon. Notably, we want to evaluate the
effects of the electrical charge ($q$), the parameter of ModMax ($\gamma $), 
$F(R)$'s parameters ($f_{R_{0}}$, and $R_{0}$) on this kind of black hole.
Our results are:

i) The effect of the electrical charge indicates that by increasing $q$, the
number of roots decreases. In other words, the higher charged black hole in $%
F(R)$-ModMax theory does not have an event horizon, and we encounter a naked
singularity (see Fig. \ref{Fig2}a).

ii) The effect of the ModMax theory reveals that the number of roots and the
radius of the event horizon increase by increasing $\gamma $. Indeed, a
black hole with large $\gamma $, has two roots (see Fig. \ref{Fig2}b).

iii) In Fig. \ref{Fig2}c, we can see the effect of $f_{R_{0}}$ on the
obtained black holes in $F(R)$-ModMax theory. This figure indicates that by
increasing $f_{R_{0}}$, the number of roots and radius of black holes
increase.

iv) The effect of $R_{0}$ appears in Fig. \ref{Fig2}d. The behavior of black
holes under this parameter is the same as the electrical charge. In other
words, by increasing $\left\vert R_{0}\right\vert $, the number of roots and
the radius of the black hole decrease.

\section{Thermodynamics}

Here, we compute the conserved and thermodynamic quantities of the
electrical charged black holes within the $F(R)$-ModMax theory. Next, we
will examine the first law of thermodynamics.

The initial step involves the computation of the Hawking temperature for the
black holes. The Hawking temperature can be determined using the following
equation 
\begin{equation}
T=\frac{\kappa }{2\pi },  \label{T}
\end{equation}%
where $\kappa $ is the superficial gravity of these black holes, which is
given by 
\begin{equation}
\kappa =\sqrt{\frac{-1}{2}\left( \nabla _{\mu }\chi _{\nu }\right) \left(
\nabla ^{\mu }\chi ^{\nu }\right) }=\left. \frac{g_{tt}^{\prime }}{2\sqrt{%
-g_{tt}g_{rr}}}=\right\vert _{r=r_{+}}=\left. \frac{g^{\prime }(r)}{2}%
\right\vert _{r=r_{+}},  \label{k}
\end{equation}%
in which $r_{+}$ and $\chi =\partial _{t}$ are the radius of the events
horizon and the Killing vector, respectively.

Before obtaining the Hawking temperature, we have to derive an expression
for the mass ($m_{0}$) using the event horizon radius ($r_{+}$), $R_{0}$,
and the charge ($q$), resulting in 
\begin{equation}
m_{0}=r_{+}-\frac{R_{0}r_{+}^{3}}{12}+\frac{q^{2}e^{-\gamma }}{\left(
1+f_{R_{0}}\right) r_{+}},  \label{mm}
\end{equation}%
where we extracted $m_{0}$ by equating $g(r)=0$.

By applying the extracted metric function (\ref{g(r)F(R)}), and by
substituting the mass (\ref{mm})\ in Eq. (\ref{k}), we get the superficial
gravity as 
\begin{equation}
\kappa =\frac{1}{2r_{+}}-\frac{R_{0}r_{+}}{8}-\frac{q^{2}e^{-\gamma }}{%
2\left( 1+f_{R_{0}}\right) r_{+}^{3}}.  \label{k2}
\end{equation}%
Now, we have reached a stage where we can acquire the Hawking temperature.
For this purpose, we replace Eq. (\ref{k2}) into Eq. (\ref{T}), which leads
to%
\begin{equation}
T=\frac{1}{4\pi r_{+}}-\frac{R_{0}r_{+}}{16\pi }-\frac{q^{2}e^{-\gamma }}{%
4\pi \left( 1+f_{R_{0}}\right) r_{+}^{3}}.  \label{TemF(R)CPMI}
\end{equation}%
As one can see, the Hawking temperature of black holes in $F(R)$-ModMax
theory is dependent on the electrical charge ($q$), the parameters of $F(R)$
gravity ($f_{R_{0}}$, and $R_{0}$), as well as ModMax's parameter ($\gamma $%
).

\begin{figure}[tbph]
\centering
\includegraphics[width=0.35\linewidth]{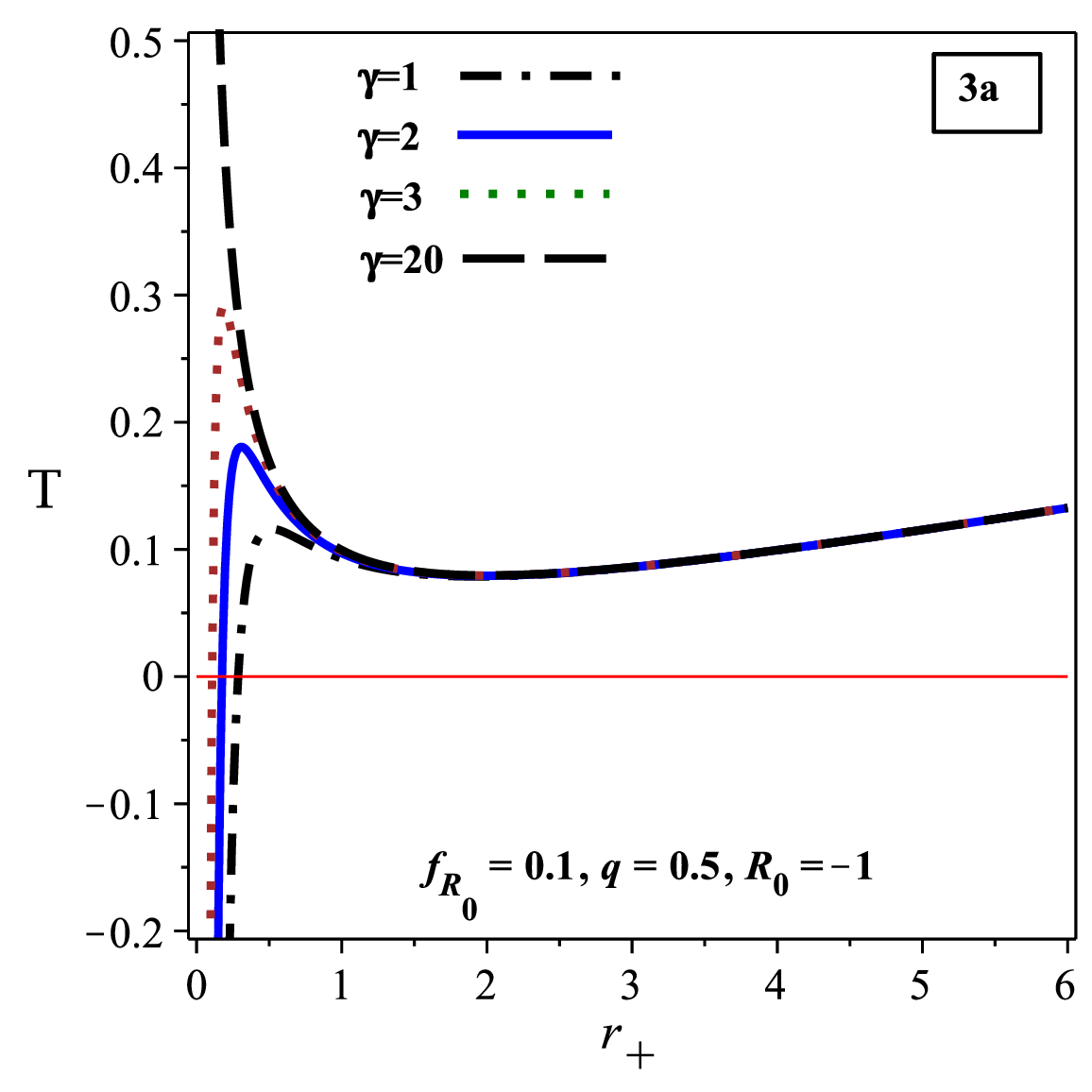} \includegraphics[width=0.35%
\linewidth]{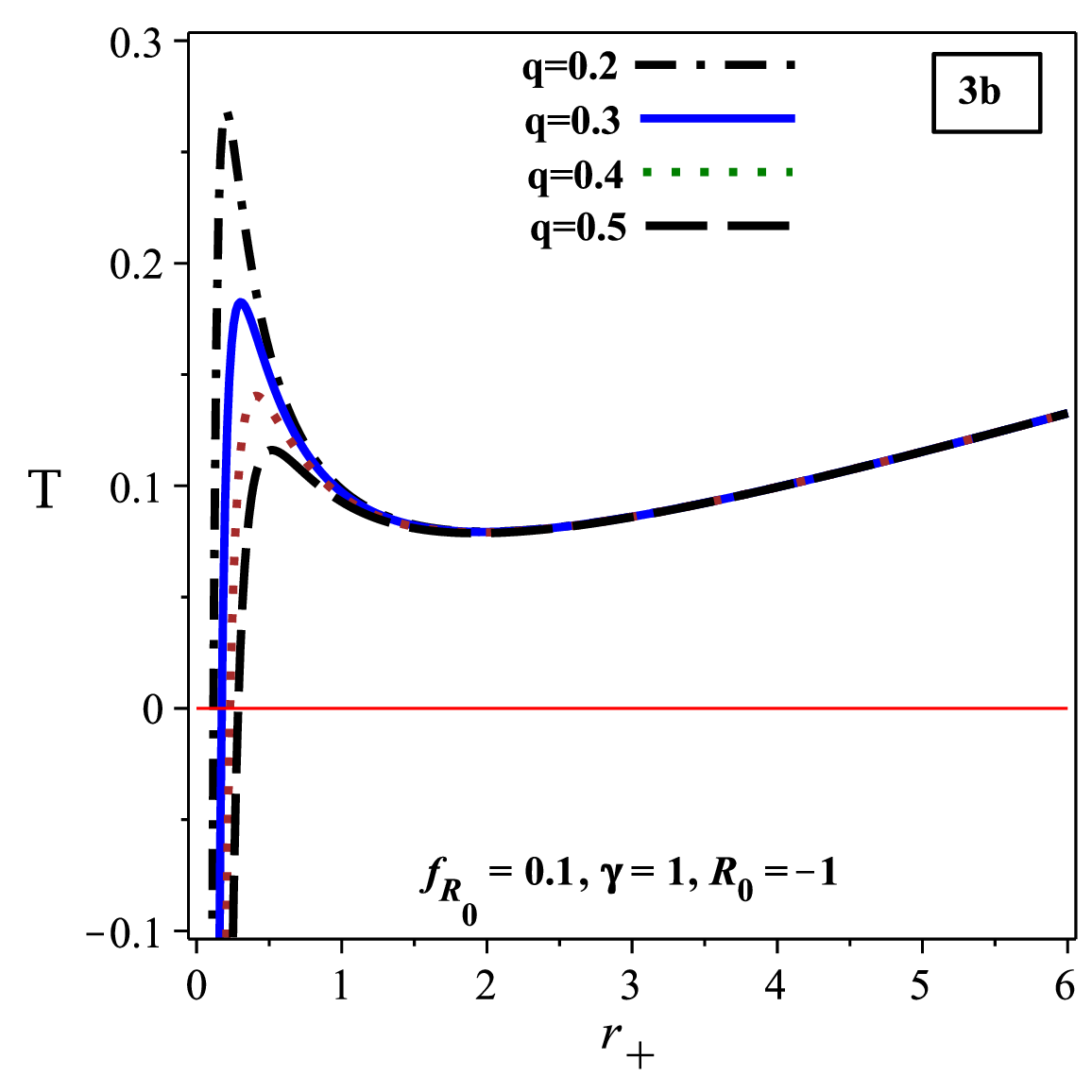}\newline
\includegraphics[width=0.35\linewidth]{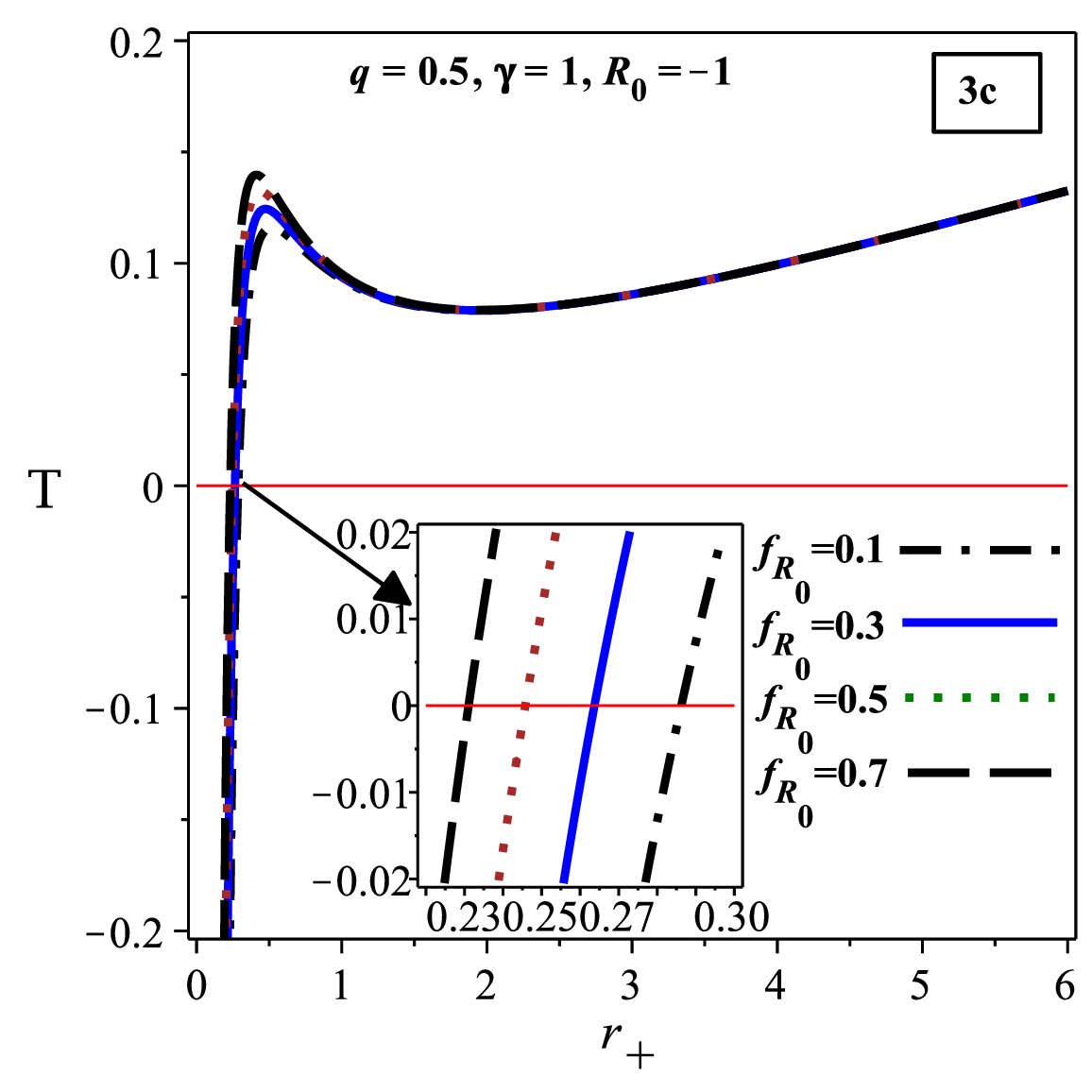} \includegraphics[width=0.35%
\linewidth]{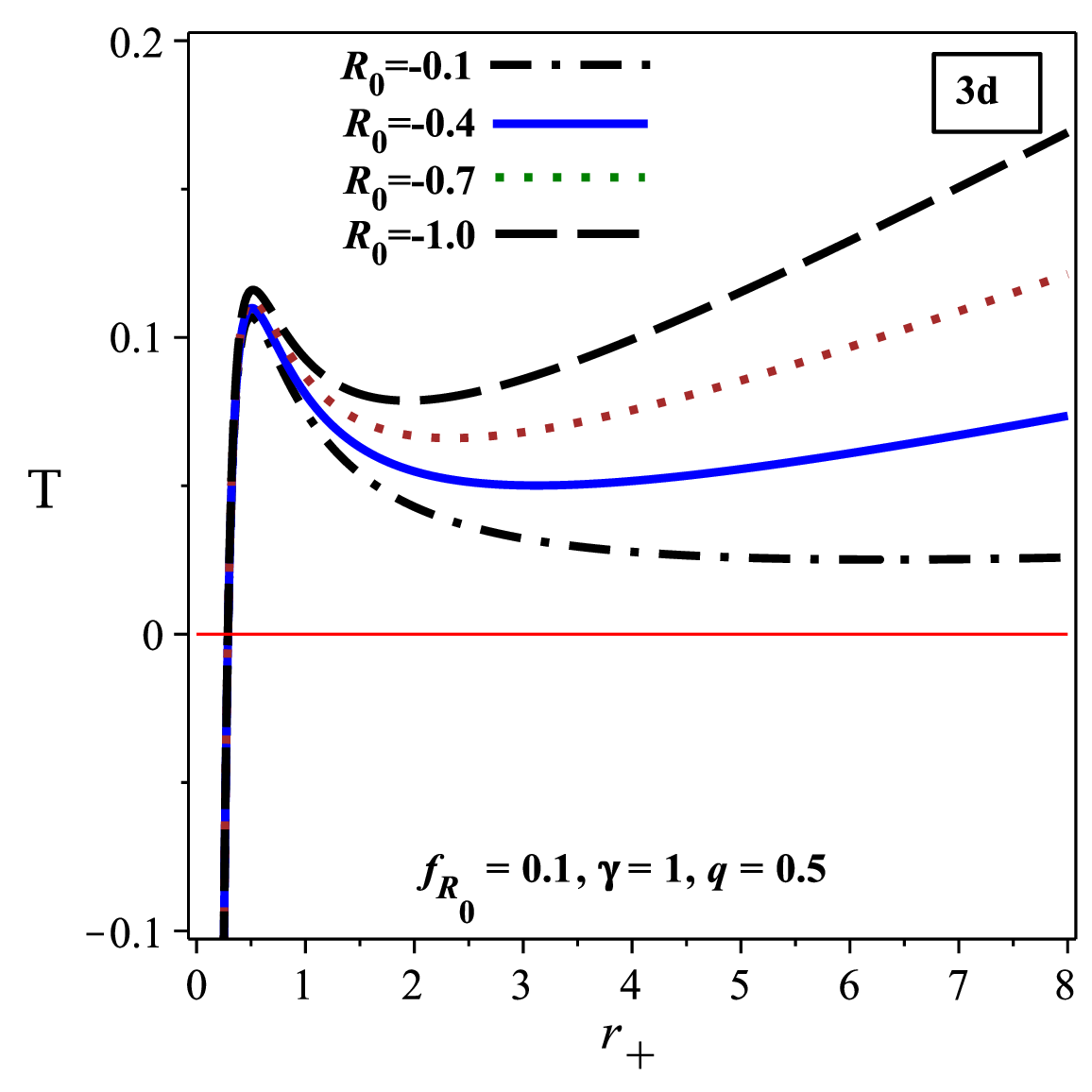}\newline
\caption{The Hawking temperature $T$ versus $r_{+}$ for different values of
the parameters.}
\label{Fig3}
\end{figure}

In classical thermodynamics, the positive (negative) values of temperature
are interpreted as (non-)physical solutions, i.e., the roots of temperature
separate physical solutions from non-physical ones. Therefore, the roots of
temperature determines bound points. To find the bound points (or real roots
of temperature), we solve the Hawking temperature. Our analysis reveals that
there is only one real root for the Hawking temperature, which is given by 
\begin{equation}
r_{+_{T=0}}=\sqrt{\frac{2}{R_{0}}\left( 1-\sqrt{1-\frac{q^{2}e^{-\gamma
}R_{0}}{1+f_{R_{0}}}}\right) }.  \label{root}
\end{equation}%
where indicates that there is no real root for the Hawking temperature when $%
\gamma \rightarrow \infty $ (see the dashed line in Fig. \ref{Fig3}a, for
more details).

We follow our study to evaluate the behavior of the high energy and
asymptotic limits of the temperature. In high energy limit of the obtained
temperature is given by 
\begin{equation}
\underset{r_{+}\rightarrow 0}{\lim }T\propto \left\{ 
\begin{array}{ccc}
\frac{1}{4\pi r_{+}}, &  & \gamma \rightarrow \infty \\ 
&  &  \\ 
-\frac{q^{2}e^{-\gamma }}{4\pi \left( 1+f_{R_{0}}\right) r_{+}^{3}}, &  & 
\text{ for small value of }\gamma%
\end{array}%
\right. ,
\end{equation}%
which reveals that the parameter of ModMax theory plays an important role in
this limit of the temperature. In other words, the temperature is always
negative for small black holes by considering the small value of $\gamma $.
But, for $\gamma \rightarrow \infty $, the temperature goes to positive
infinity (see four diagrams in Fig. \ref{Fig3}a).

On the other hand, the asymptotic limit of the temperature only depends on $%
R_{0}$, i.e. 
\begin{equation}
\underset{r_{+}\rightarrow \infty }{\lim }T\propto -\frac{R_{0}r_{+}}{16\pi }%
,  \label{asymp}
\end{equation}%
which is dependent on one of the parameters of $F(R)$ gravity. The
asymptotic limit of the temperature is always positive (negative) when $%
R_{0}<0$ ($R_{0}>0$).

To confirm the obtained results of the various behaviors (high energy and
asymptotic limit) for the Hawking temperature and study other effects of the
mentioned parameters such as $q$, and $f_{R_{0}}$, we plot the Hawking
temperature versus $r_{+}$ in Fig. \ref{Fig3}.

We can see the effects of different parameters on the Hawking temperature in
Fig. \ref{Fig3}. There are two extrema points that belong to the maximum and
minimum values of the temperature. Actually, by increasing the radius of the
black hole, the temperature reaches a maximum value (the first extremum) and
then decreases to a minimum value (the second extremum). After the second
extremum point, the temperature increases.

Our analysis of the effects of $\gamma $, $q$, $f_{R_{0}}$, and $R_{0}$ on $%
r_{+_{T=0}}$ (root of the temperature) reveal that:

i) $r_{+_{T=0}}$ decreases by increasing the parameter of ModMax theory, and
finally, for the very large value of $\gamma $, we have no root, as we
expected from the equation (\ref{root}). See\ Fig. \ref{Fig3}a, for more
details.

ii) The effect of the electrical charge on the root of the temperature shows
that by increasing $q$, $r_{+_{T=0}}$ increases (see Fig. \ref{Fig3}b).

iii) Fig. \ref{Fig3}c indicates that $r_{+_{T=0}}$ decreases by increasing $%
f_{R_{0}}$.

iv) The root of the temperature is not very sensitive to changes of $R_{0}$
(see Fig. \ref{Fig3}d). But the asymptotic limit of the temperature depends
on completely this parameter, as we expected it from Eq. (\ref{asymp}).

Our results reveal that the small black holes\ (i.e., $r_{+}<r_{+_{T=0}}$),
cannot be physical because the temperature is negative in this area, except
for the large value of $\gamma $. In other words, small black holes are
physical, provided the parameter of ModMax theory has a large value.

We can get the electric charge of the black hole in $F(R)$-ModMax theory, by
using the Gauss law, which leads to 
\begin{equation}
Q=q.  \label{Q}
\end{equation}

One can find the electric potential at the event horizon ($U$) in the
following form 
\begin{equation}
U=-\int_{r_{+}}^{+\infty }F_{tr}dr=\frac{qe^{-\gamma }}{r_{+}}.
\label{elcpoF(R)CPMI}
\end{equation}%
where $F_{tr}=\frac{qe^{-\gamma }}{r^{2}}$.

We can apply a modification of the area law in the $F(R)$ theory of gravity 
\cite{Cognola2005}, to extract the entropy of black holes, which yields 
\begin{equation}
S=\frac{A(1+f_{R_{0}})}{4},  \label{SFR}
\end{equation}%
in which $A$ is the horizon area as 
\begin{equation}
A=\left. \int_{0}^{2\pi }\int_{0}^{\pi }\sqrt{g_{\theta \theta }g_{\varphi
\varphi }}\right\vert _{r=r_{+}}=\left. 4\pi r^{2}\right\vert
_{r=r_{+}}=4\pi r_{+}^{2}.  \label{A}
\end{equation}%
Now, we can obtain the entropy of ModMax-black holes in $F(R)$ gravity by
replacing Eq. (\ref{A}) within Eq. (\ref{SFR}), which leads to 
\begin{equation}
S=\pi (1+f_{R_{0}})r_{+}^{2}.  \label{S}
\end{equation}

The Ashtekar-Magnon-Das (AMD) approach enables us to determine the total
mass of black holes in the $F(R)$-ModMax theory \cite{AMDI,AMDII}, in the
following form 
\begin{equation}
M=\frac{m_{0}\left( 1+f_{R_{0}}\right) }{2}.  \label{AMDMass}
\end{equation}%
Here, we expand the total mass by substituting the mass (\ref{mm}) within
the equation (\ref{AMDMass}), and we get 
\begin{equation}
M=\frac{\left( 1+f_{R_{0}}\right) r_{+}}{2}\left( 1-\frac{R_{0}r_{+}^{2}}{12}%
\right) +\frac{q^{2}e^{-\gamma }}{2r_{+}},  \label{MM}
\end{equation}%
where indicates that the total mass is dependent on the parameters of the
electrical charge, the parameters of $F(R)$ gravity, as well as ModMax's
parameter.

\begin{figure}[tbph]
\centering
\includegraphics[width=0.35\linewidth]{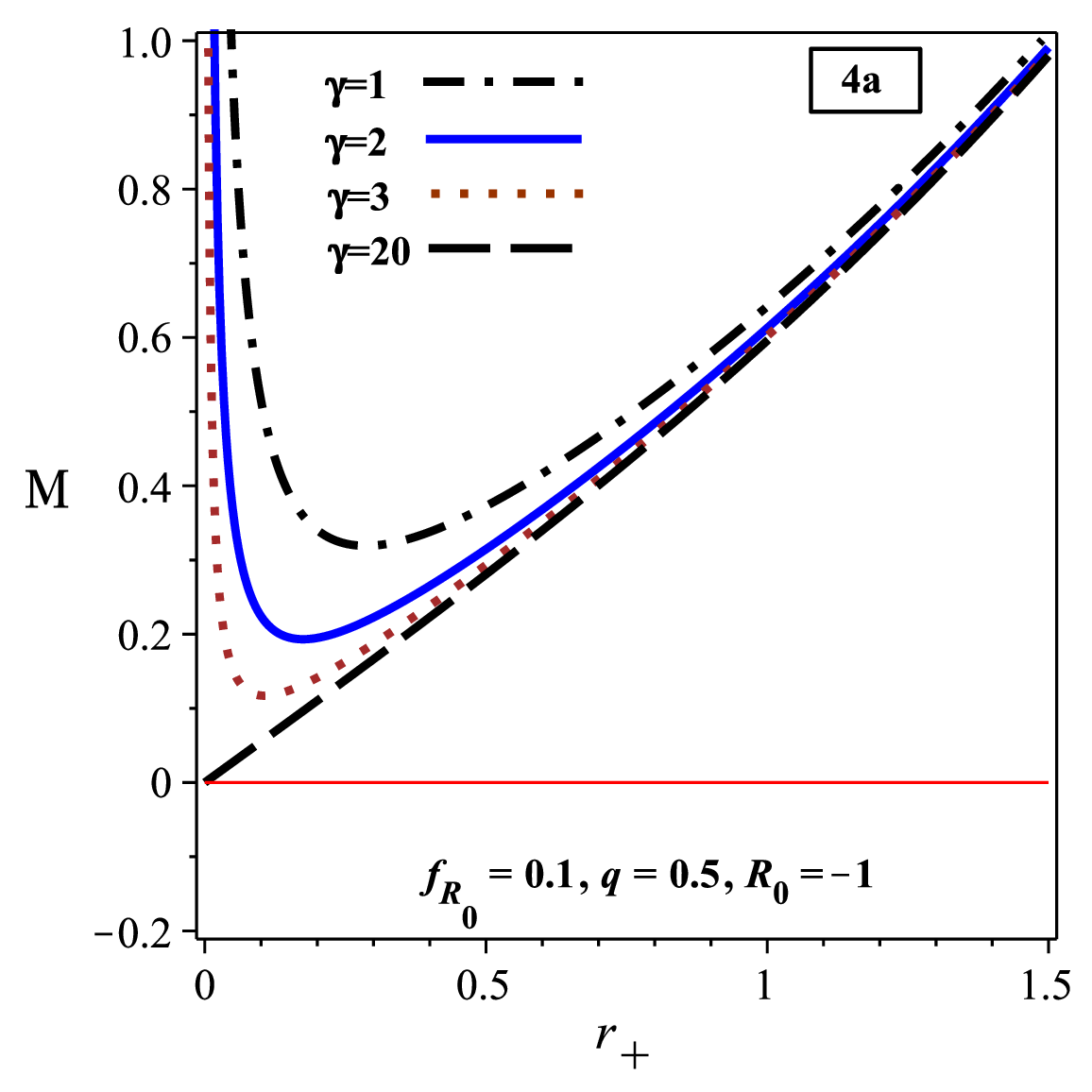} \includegraphics[width=0.35%
\linewidth]{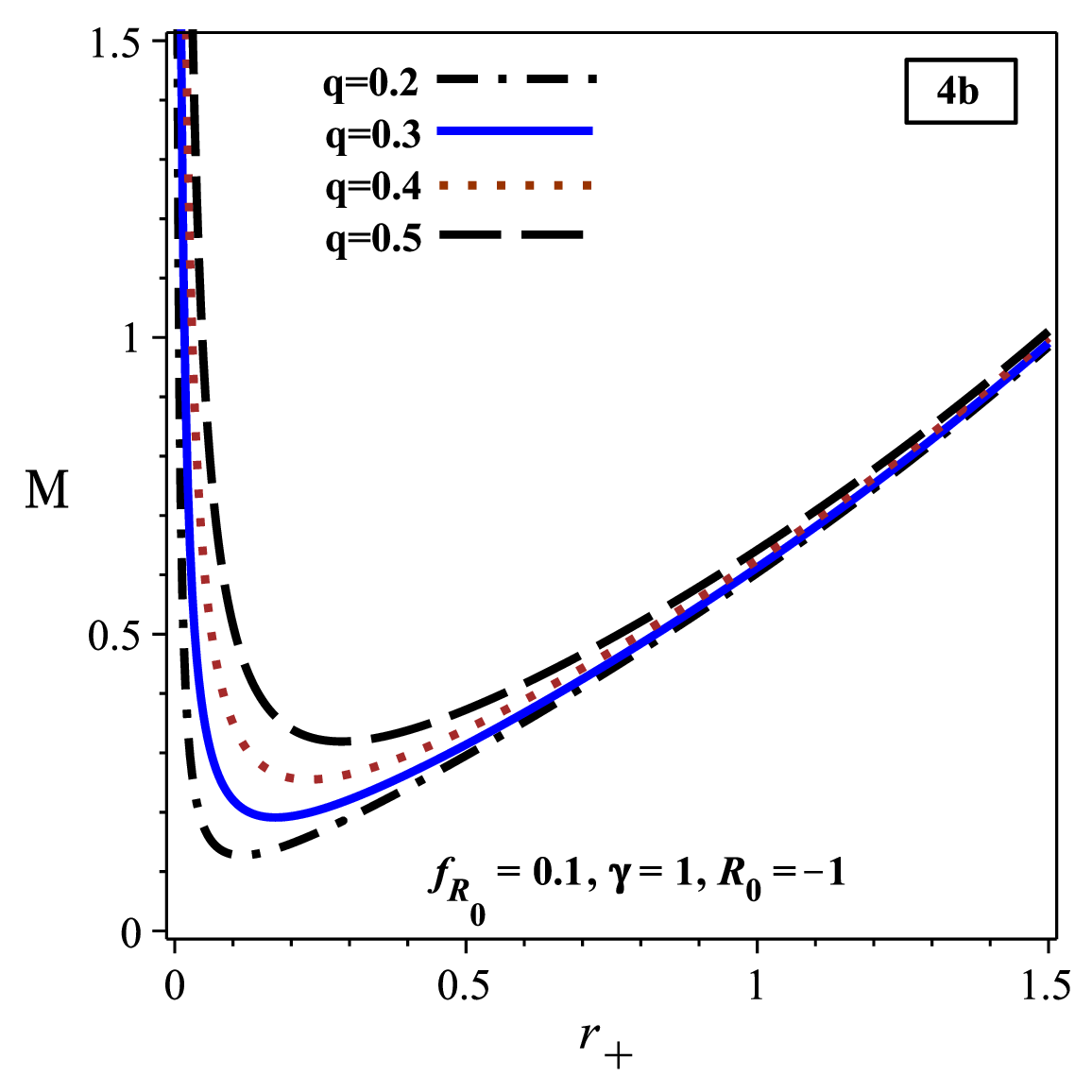}\newline
\includegraphics[width=0.35\linewidth]{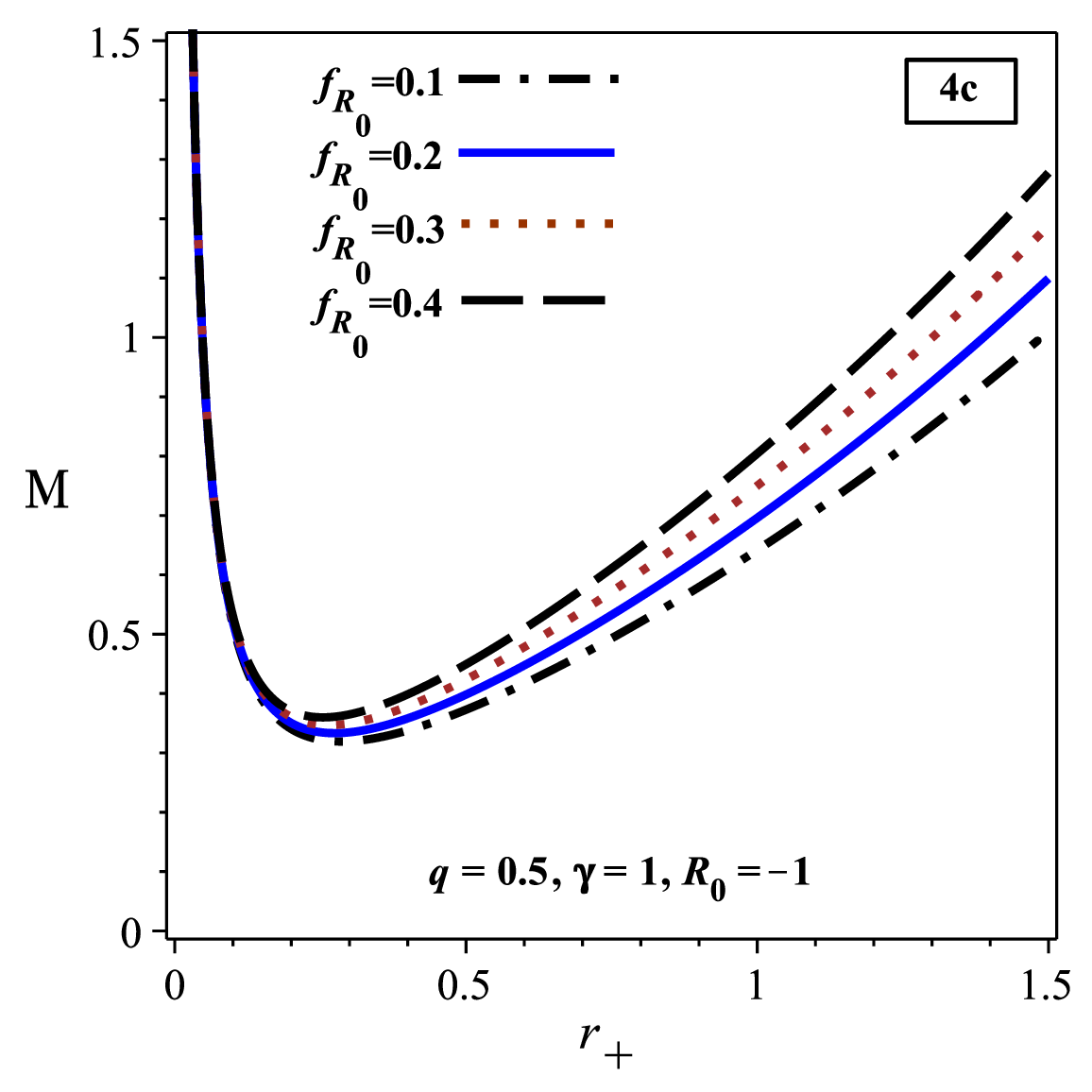} \includegraphics[width=0.35%
\linewidth]{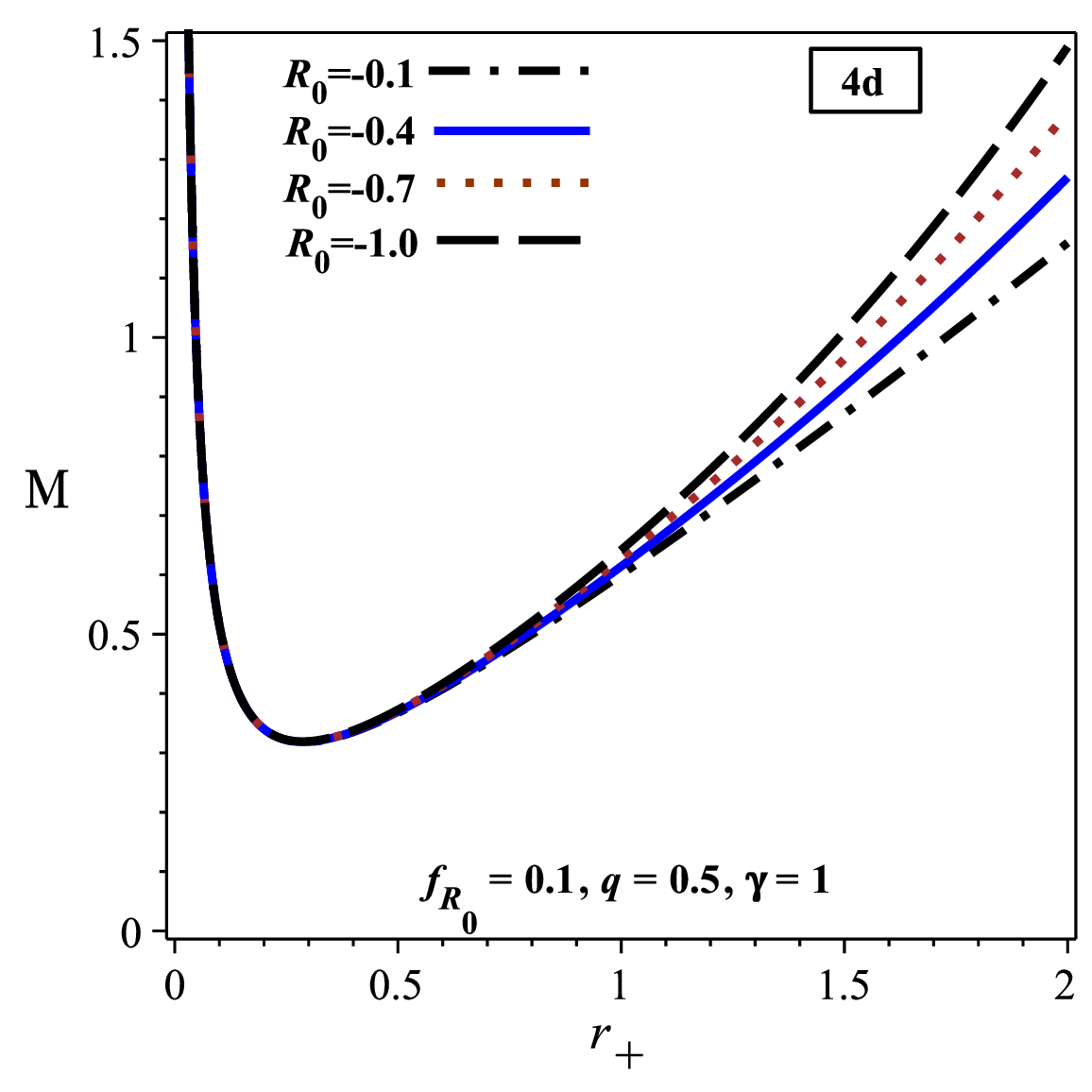}\newline
\caption{The total mass $M$ versus $r_{+}$ for different parameters.}
\label{Fig4}
\end{figure}

In high energy limit of the total mass is given by 
\begin{equation}
\underset{r_{+}\rightarrow 0}{\lim }M\propto \frac{q^{2}e^{-\gamma }}{2r_{+}}%
,
\end{equation}%
which depends on $q$ and $\gamma $. The total mass of small black holes is
always positive for finite values of $\gamma $. Also, $M$ of small black
holes is zero (i.e., $\underset{r_{+}\rightarrow 0}{\lim }M=0$) when $%
\gamma\rightarrow \infty $, see the dashed line in Fig. \ref{Fig4}a.

The asymptotic limit of $M$ is obtained 
\begin{equation}
\underset{r_{+}\rightarrow \infty }{\lim }M\propto -\frac{\left(
1+f_{R_{0}}\right) R_{0}r_{+}^{3}}{24},  \label{asypM}
\end{equation}%
which depends on the parameters of $F(R)$ gravity ($f_{R_{0}}$, and $R_{0}$%
). Also, the asymptotic limit of $M$ is always positive (negative) when $%
R_{0}<0$ ($R_{0}>0$).

To see the effects of $\gamma $, $q$, $f_{R_{0}}$, and $R_{0}$ on the total
mass of black holes, we plot Fig. \ref{Fig4}. Our analysis states that for a
finite value of $\gamma $, there is an extreme point (a minimum point) of
the total mass. Indeed, $M$ decreases by increasing $r_{+}$, and reaches a
minimum value. After this extreme point, the total mass increases by
increasing the radius of black holes. Notably, there is no extreme point for
the total mass when we consider the large value of $\gamma $ (see Fig. \ref%
{Fig4}a, for more details). In this case, the total mass is an increasing
function of $r_{+}$.

In summary, our examination of the impacts of different factors on the total
mass are as follows:

i) The extreme point depends on the parameter of ModMax theory. Considering
a finite value of ModMax's parameter, by increasing $\gamma $, the extreme
point decreases and the minimum point is located at a smaller radius (see
Fig. \ref{Fig4}a).

ii) The effect of electrical charge on the total mass presented in Fig. \ref%
{Fig4}b. The results indicate that by increasing $q$, the extreme point
increases, and it is located at a large radius.

iii) The minimum of total mass increases by increasing $f_{R_{0}}$. In
addition, for the large values of $f_{R_{0}}$, the extreme point is located
in a smaller radius (see Fig. \ref{Fig4}c).

iv) We can see the effect of $R_{0}$ on the total mass in Fig. \ref{Fig4}d.
The extreme point is not very sensitive to the parameter of $R_{0}$. As one
can see in Eq. (\ref{asypM}) and Fig. \ref{Fig4}d, the asymptotic limit of $%
M $ is sensitive to $R_{0}$.

Now, we can evaluate the first law of thermodynamics. In other words, the
obtained conserved and thermodynamics quantities in Eqs. (\ref{TemF(R)CPMI}%
), (\ref{Q}), (\ref{elcpoF(R)CPMI}), (\ref{S}), and (\ref{MM}), satisfy the
first law of thermodynamics in the following format 
\begin{equation}
dM=TdS+UdQ,
\end{equation}%
where $T=\left( \frac{\partial M}{\partial S}\right) _{Q}$, and $U=\left( 
\frac{\partial M}{\partial Q}\right) _{S}$, respectively, are in agreement
with the obtained relations in Eqs. (\ref{TemF(R)CPMI}) and (\ref%
{elcpoF(R)CPMI}).

\section{Thermal Stability}

In order to investigate the thermal stability of a black hole as a
thermodynamic system, our focus lies on examining the impact of parameters
within the $F(R)$-ModMax theory. This analysis will be conducted through the
utilization of both heat capacity and Helmholtz free energy.

\subsection{Heat Capacity}

In the domain of the canonical ensemble, the local stability of a
thermodynamic system can be assessed through the utilization of heat
capacity. The heat capacity, whether positive or negative, serves as an
indicator of the system's thermal stability. It is worth noting that a
positive heat capacity corresponds to thermal stability. Therefore, we
examine the local stability of black holes in the $F(R)$-ModMax theory by
employing the heat capacity.

The heat capacity is defined in the following form 
\begin{equation}
C_{Q}=\frac{T}{\left( \frac{\partial T}{\partial S}\right) _{Q}}=\frac{%
\left( \frac{\partial M\left( S,Q\right) }{\partial \widetilde{S}}\right)
_{Q}}{\left( \frac{\partial ^{2}M\left( S,Q\right) }{\partial S^{2}}\right)
_{Q}},  \label{Heat}
\end{equation}%
to obtain the heat capacity, we re-write the total mass and the Hawking
temperature of the black hole (\ref{MM}) in terms of the electrical charge (%
\ref{Q}), and the entropy (\ref{S}), which lead to 
\begin{eqnarray}
M\left( S,Q\right) &=&\frac{\left( S+\pi Q^{2}e^{-\gamma }\right) \left(
1+f_{R_{0}}\right) -\frac{R_{0}S^{2}}{12\pi }}{\sqrt{\pi S\left(
1+f_{R_{0}}\right) }},  \label{MSQ} \\
&&  \notag \\
T &=&\left( \frac{\partial M\left( S,Q\right) }{\partial S}\right) _{Q}=%
\frac{\left( S-\pi Q^{2}e^{-\gamma }\right) \left( 1+f_{R_{0}}\right) -\frac{%
R_{0}S^{2}}{\pi }}{4S\sqrt{\pi S\left( 1+f_{R_{0}}\right) }}.  \label{TM}
\end{eqnarray}

Now, we can get the heat capacity by considering Eqs. (\ref{MSQ}) and (\ref%
{TM}) within Eq. (\ref{Heat}), and after some calculation, we have 
\begin{equation}
C_{Q}=\frac{2S\left[ \left( S-\pi Q^{2}e^{-\gamma }\right) \left(
1+f_{R_{0}}\right) -\frac{R_{0}S^{2}}{4\pi }\right] }{4\left( 3\pi
Q^{2}e^{-\gamma }-S\right) \left( 1+f_{R_{0}}\right) -\frac{R_{0}S^{2}}{4\pi 
}}.  \label{Heat1}
\end{equation}

In the realm of black holes, the root of heat capacity $\left(
C_{Q}=T=0\right) $ is considered as a border line between non-physical $%
\left( T<0\right) $, and physical $\left( T>0\right) $ black holes.
Hereafter, we name the root of heat capacity (or the root of temperature) as
a physical limitation point. In simpler terms, the heat capacity undergoes a
sign change at the point of physical limitation. Furthermore, it has been
suggested that the divergences in heat capacity serve as crucial indicators
of phase transition critical points in black holes. Consequently, the heat
capacity can be utilized to identify both the phase transition critical and
physical limitation points in black holes. So, we determine these points by
solving the following relations 
\begin{equation}
\left\{ 
\begin{array}{ccc}
T=\left( \frac{\partial M\left( S,Q\right) }{\partial S}\right) _{Q}=0, &  & 
\text{physical limitation points} \\ 
&  &  \\ 
\left( \frac{\partial ^{2}M\left( S,Q\right) }{\partial S^{2}}\right) _{Q}=0
&  & \text{phase transition critical points}%
\end{array}%
\right. .  \label{PhysBound}
\end{equation}

The physical limitation point is obtained by utilizing Equation (\ref{TM})
and solving it with respect to entropy, resulting in 
\begin{equation}
S_{root}=\frac{2\pi \left( 1+f_{R_{0}}\right) }{R_{0}}\left[ 1-\sqrt{1-\frac{%
R_{0}Q^{2}e^{-\gamma }}{1+f_{R_{0}}}}\right] .  \label{RootT}
\end{equation}
The above relation imposes a constraint on $R_{0}$. Indeed, to have the real
root, it is necessary to adhere to the condition $R_{0}\leq\frac{1+f_{R_{0}}%
}{Q^{2}e^{-\gamma }}$. In addition, the root of the obtained temperature (%
\ref{RootT}), depends on the parameters of $\gamma $, $Q$, $f_{R_{0}}$, and $%
R_{0}$. It is worth mentioning that in the previous section, we deliberated
on the impact of these parameters on the temperature's root.

To evaluate the phase transition critical points, we have to solve the
relation $\left( \frac{\partial ^{2}M\left( S,Q\right) }{\partial S^{2}}%
\right) _{Q}=0$. We get two divergence points, which are%
\begin{equation}
\left\{ 
\begin{array}{c}
S_{div_{1}}=\frac{-2\pi \left( 1+f_{R_{0}}\right) }{R_{0}}\left( 1-\sqrt{1+%
\frac{3R_{0}Q^{2}e^{-\gamma }}{1+f_{R_{0}}}}\right) \\ 
\\ 
S_{div_{2}}=\frac{-2\pi \left( 1+f_{R_{0}}\right) }{R_{0}}\left( 1+\sqrt{1+%
\frac{3R_{0}Q^{2}e^{-\gamma }}{1+f_{R_{0}}}}\right)%
\end{array}%
\right. ,  \label{rdivHeat}
\end{equation}%
which impose the constraint $R_{0}\geq \frac{-\left(1+f_{R_{0}}\right) }{%
3Q^{2}e^{-\gamma }}$, in order to have the real divergent point(s). Our
analysis from Eq. (\ref{rdivHeat}) reveals that the heat capacity may have
three different cases:

The first case: there is one divergence point when $\gamma \rightarrow
\infty $. Indeed, the real and positive divergence point is located at $%
S_{div_{2}}=\frac{-4\pi \left( 1+f_{R_{0}}\right) }{R_{0}}$, when $R_{0}<0$.

The second case: by considering $R_{0}=\frac{-\left( 1+f_{R_{0}}\right) }{%
3Q^{2}e^{-\gamma }}$, the divergence points reduce from two points to one
point. In other words, we encounter with one divergence point at $%
S_{div_{1}}=S_{div_{2}}=\frac{-2\pi \left( 1+f_{R_{0}}\right) }{R_{0}}$,
when $R_{0}=\frac{-\left( 1+f_{R_{0}}\right) }{3Q^{2}e^{-\gamma }}$.

The third case: for finite value of $\gamma $, and $R_{0}\geq \frac{-\left(
1+f_{R_{0}}\right) }{3Q^{2}e^{-\gamma }}$, there are two divergence points
which are located in $S_{div_{1}}$, and $S_{div_{2}}$.

\begin{figure}[tbph]
\centering
\includegraphics[width=0.35\linewidth]{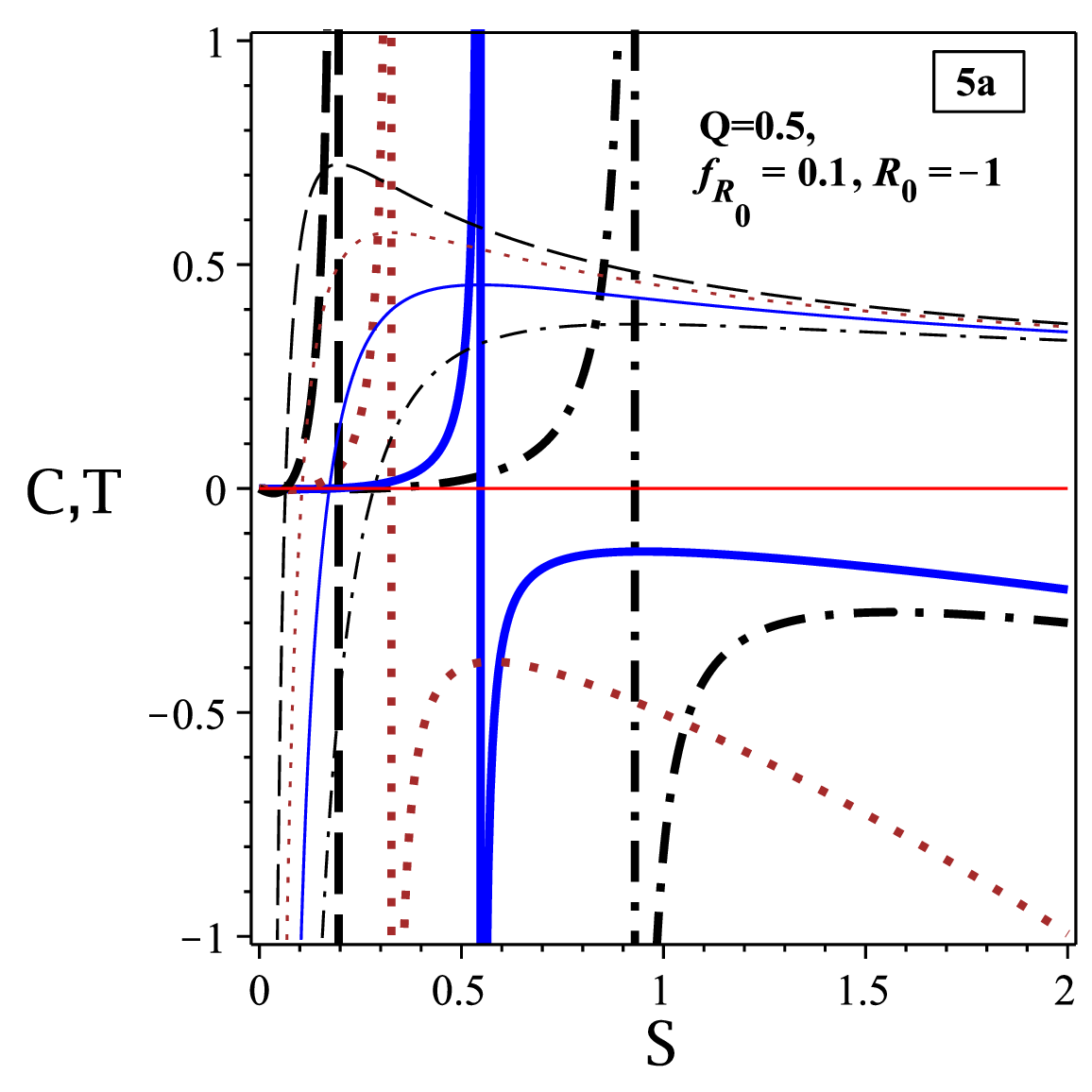} \includegraphics[width=0.35%
\linewidth]{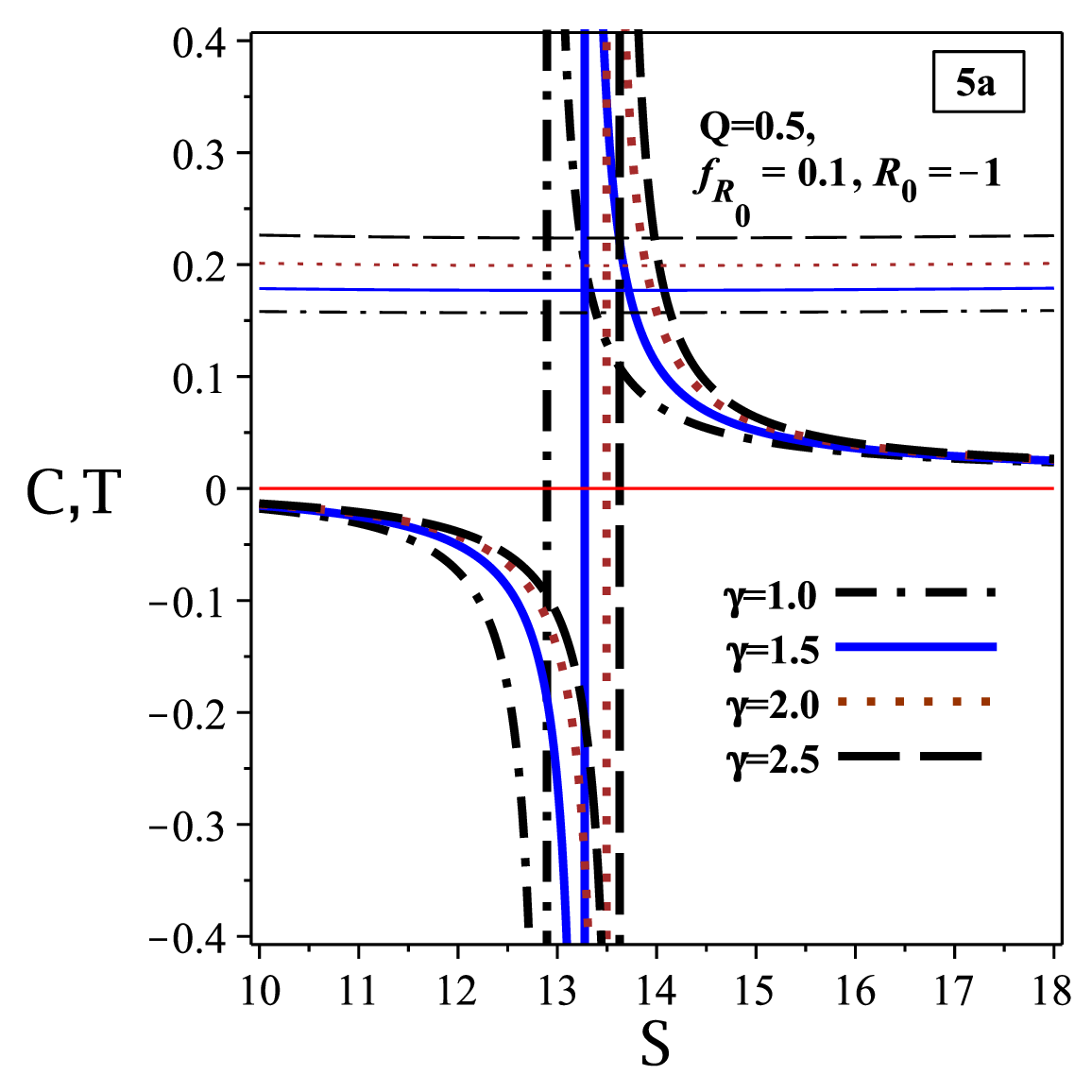}\newline
\includegraphics[width=0.35\linewidth]{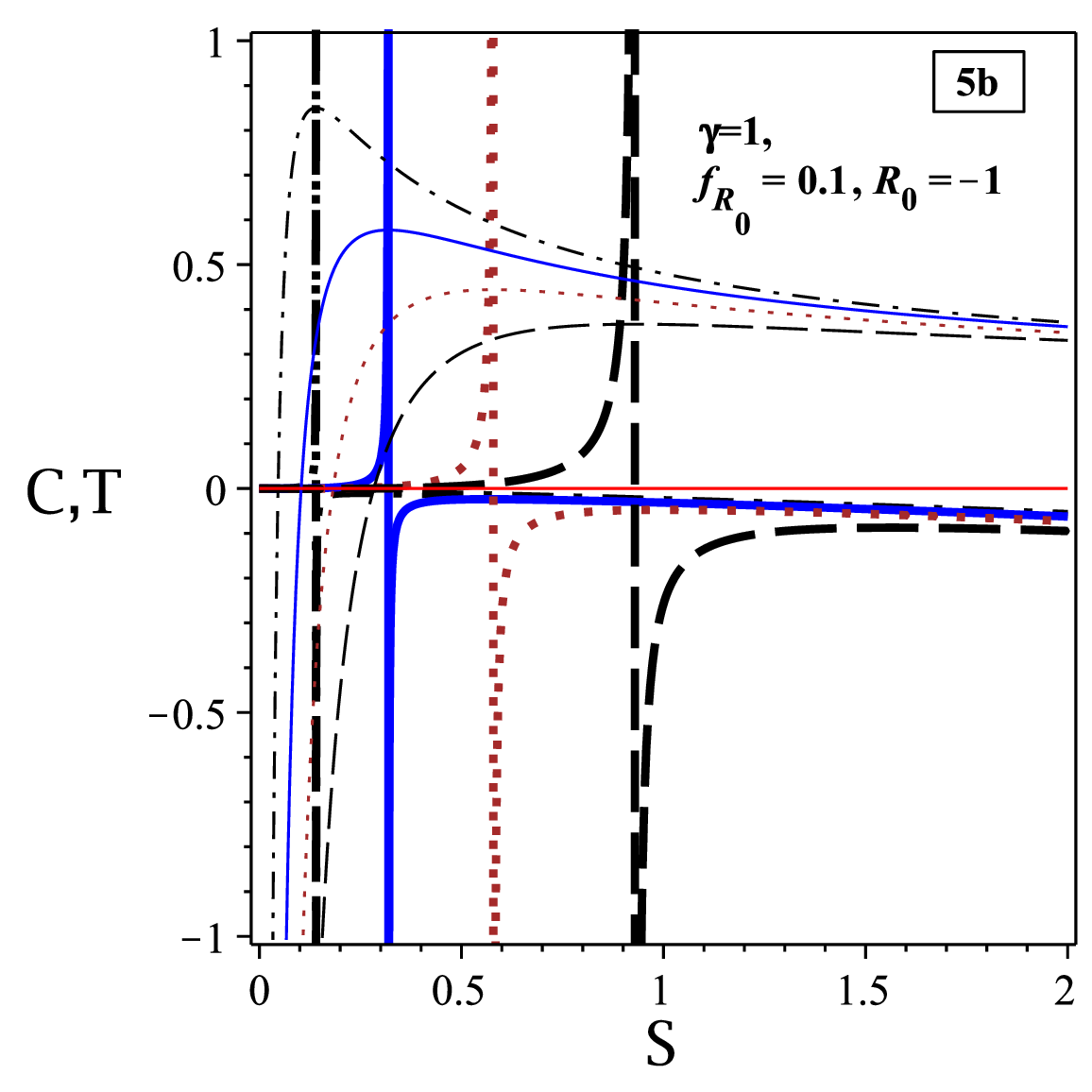} \includegraphics[width=0.35%
\linewidth]{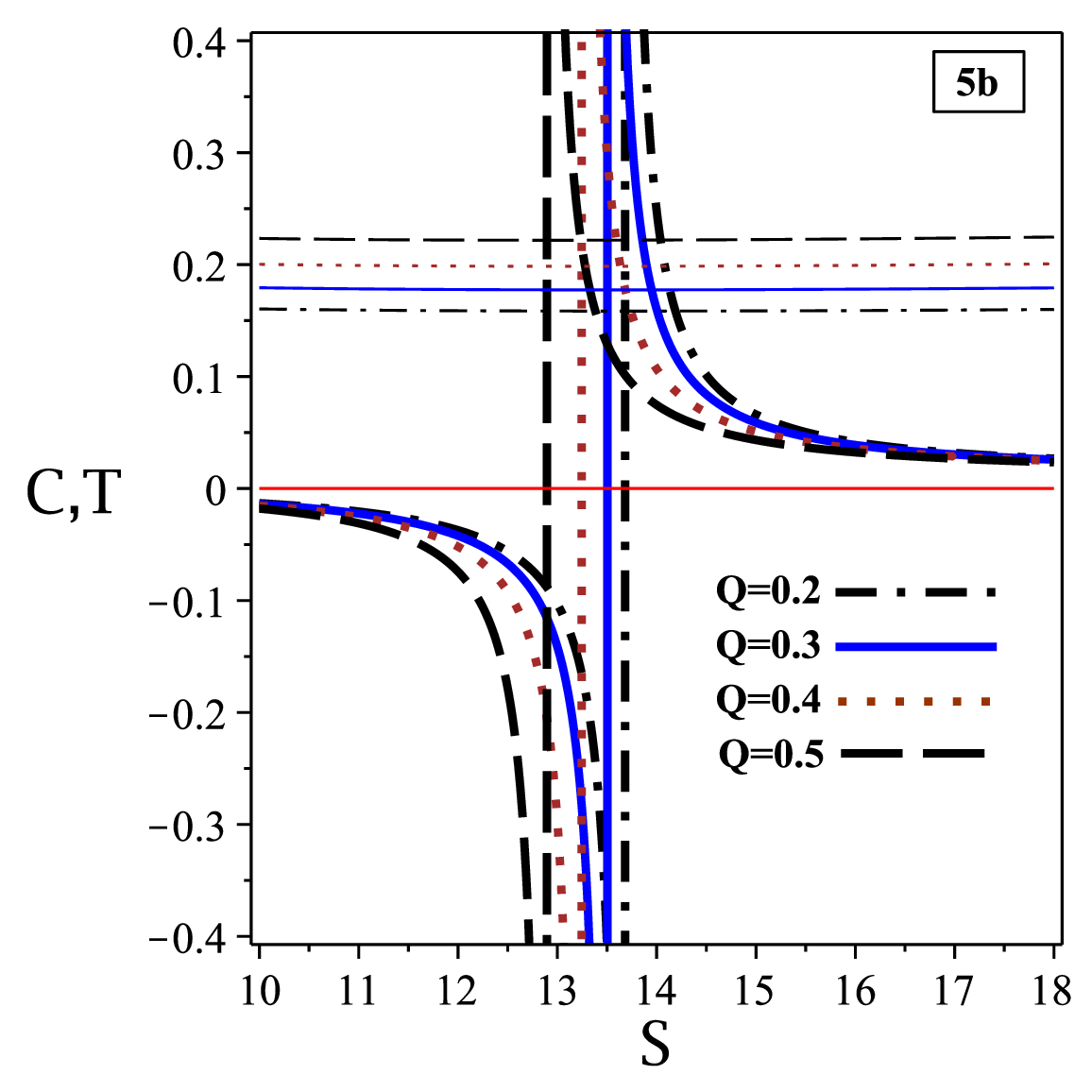}\newline
\includegraphics[width=0.35\linewidth]{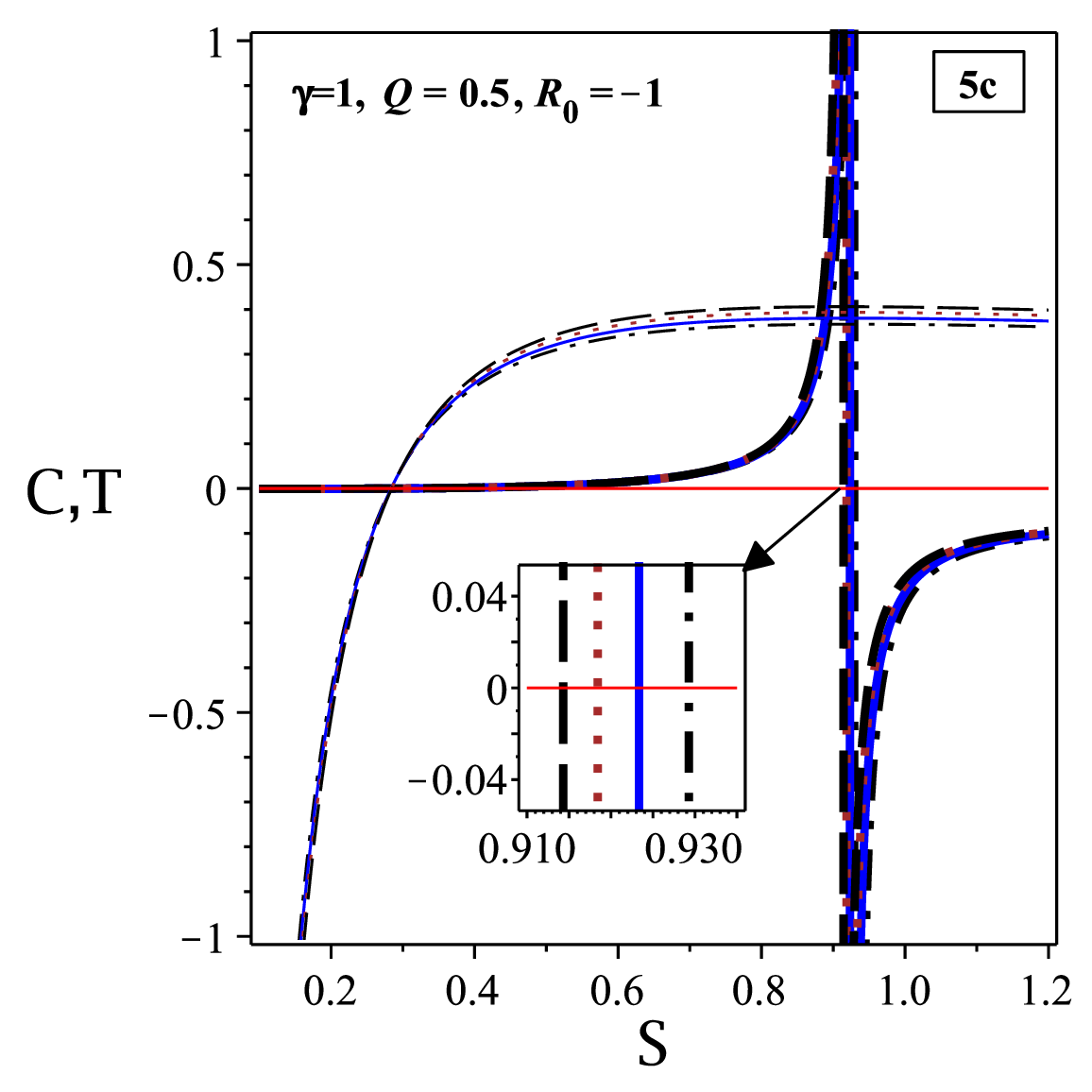} \includegraphics[width=0.35%
\linewidth]{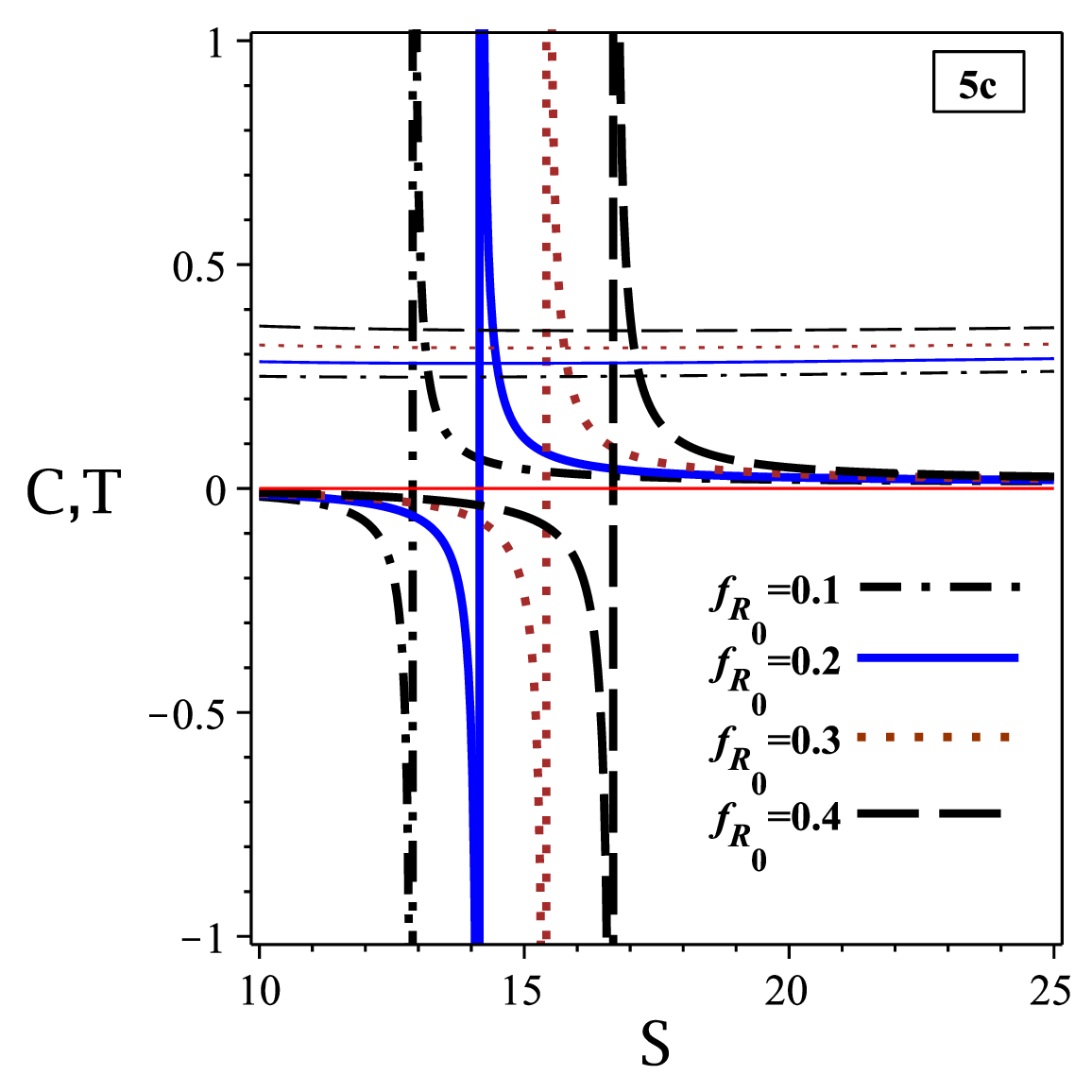}\newline
\includegraphics[width=0.35\linewidth]{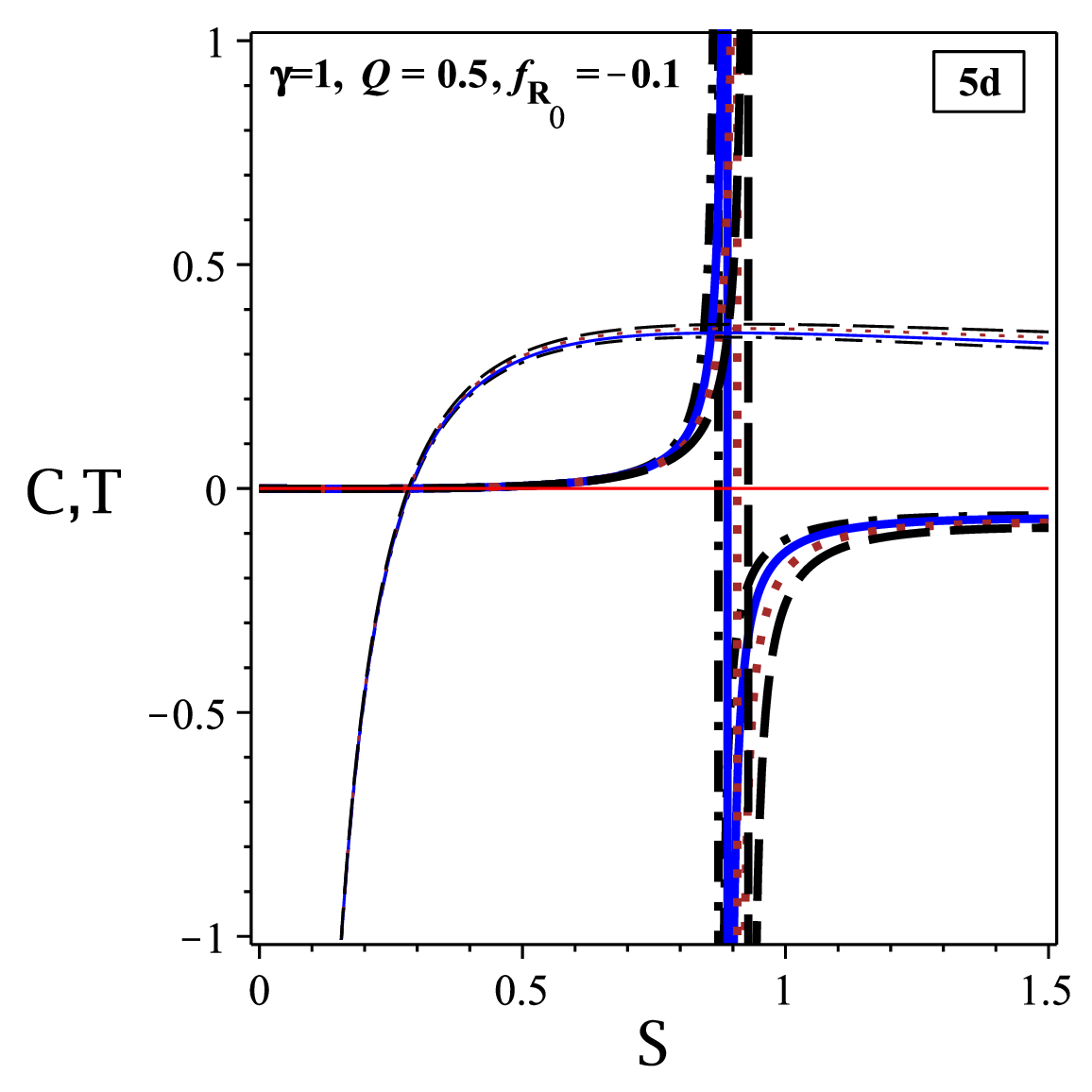} \includegraphics[width=0.35%
\linewidth]{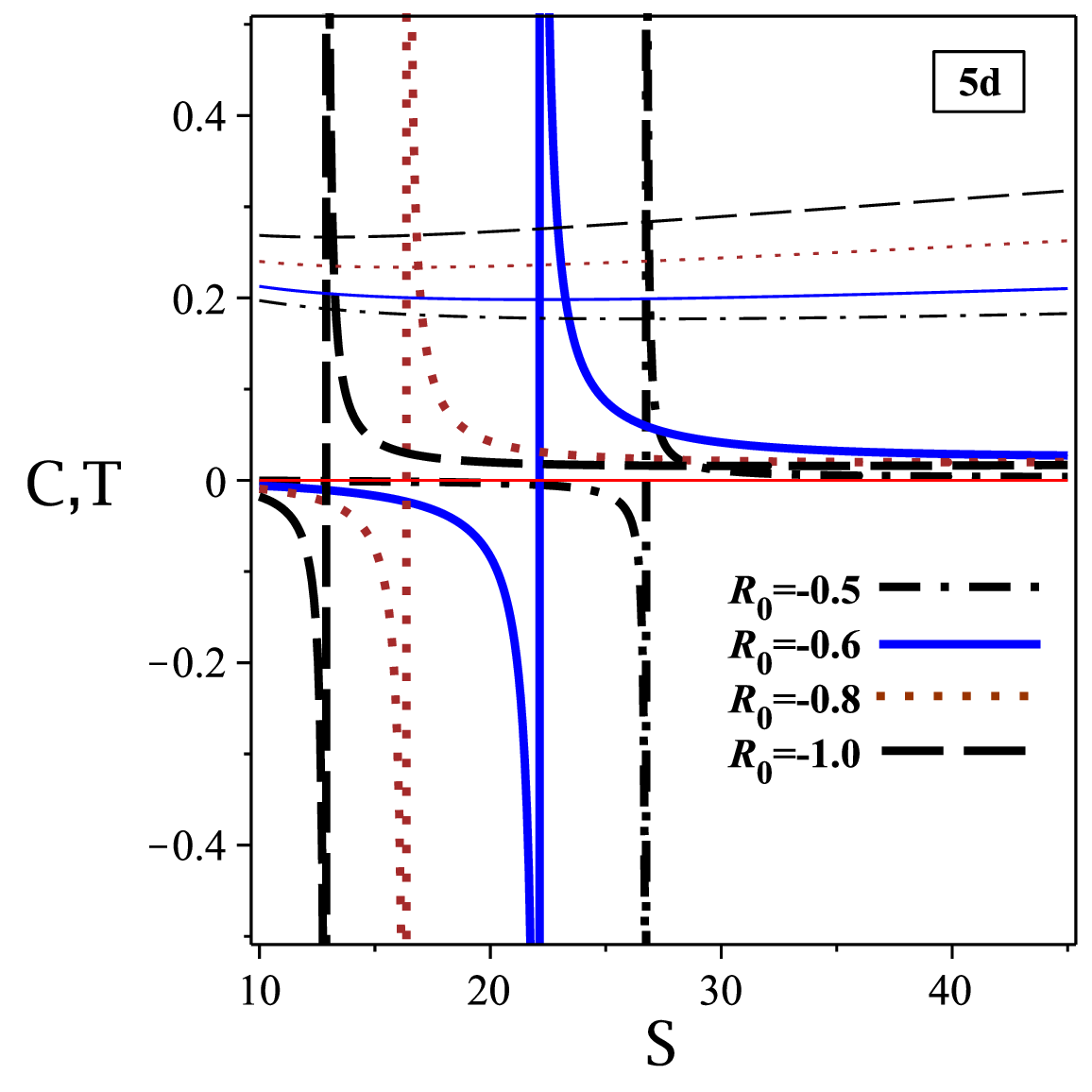}\newline
\caption{The heat capacity (bold lines) $C$ and temperature (thin lines) $T$
versus $S$ for different values of the parameters. Note: the left and right
panels are related to a case, which is plotted in different scales.}
\label{Fig5}
\end{figure}

Now we are in a position to study the local stability by using the obtained
temperature and heat capacity. Our findings are given with more details in
Fig. \ref{Fig5} and Table. \ref{tab1}.

\begin{table}[tbp]
\caption{The local stability of the black holes in $F(R)$-MadMax theory for
finite value of $\protect\gamma$, and $R_{0}<0$.}
\label{tab1}
\begin{center}
\begin{tabular}{|c|c|c|}
\hline
$T>0$ & $C_{Q}>0$ & local stability and physical area \\ \hline\hline
$S>S_{root}$ & $%
\begin{array}{c}
S_{root}<S<S_{div_{1}} \\ 
S>S_{div_{2}}%
\end{array}%
$ & $%
\begin{array}{c}
S_{root}<S<S_{div_{1}} \\ 
S>S_{div_{2}}%
\end{array}%
$ \\ \hline
\end{tabular}%
\end{center}
\end{table}

Our study of the temperature and the heat capacity, simultaneously, reveals
some information about the physical and local stability of the black holes
in $F(R)$-ModMax theory. We mention the which are:

i) There are two physical and stable areas. These areas are located at $%
S_{root}<S<S_{div_{1}}$, and $S>S_{div_{2}}$. Indeed, the temperature and
the heat capacity are positive in these areas. This result also indicates
that the black holes in the ranges $S<S_{root}$, and $%
S_{div_{1}}<S<S_{div_{2}}$, cannot be physical and stable objects, whereas
the medium black holes ($S_{root}<S<S_{div_{1}}$), and large black holes ($%
S>S_{div_{2}}$) satisfy the local stability and are physical and stable
objects. See Fig. \ref{Fig5}, and Table. \ref{tab1}, for more details.

ii) By increasing the parameter of ModMax theory, $S_{root}$, and $%
S_{div_{1}}$ decrease but $S_{div_{2}}$ increases. In other words, although
the physical area increases, but the local stability for black holes in the
ranges $S_{root}<S<S_{div_{1}}$, and $S>S_{div_{2}}$ decreases by increasing 
$\gamma $ (see two panels in Fig. \ref{Fig5}a). In addition, in the limit $%
\gamma \rightarrow \infty $, there is only one divergence point, i.e., $%
S_{div_{2}}$, which states that the large black holes (i.e., $S>S_{div_{2}}$%
) can satisfy physical and local stability, simultaneously.

iii) We can see the effect of the electrical charge on physical and local
stability in two panels of Fig. \ref{Fig5}b. The results show that by
increasing $Q$, $S_{root}$, and $S_{div_{1}}$ increase but $S_{div_{2}}$
decreases. It means that the local stability areas (i.e., $%
S_{root}<S<S_{div_{1}}$, and $S>S_{div_{2}}$) increase by increasing $Q$. In
other words, the higher electrically charged black holes have large physical
and stability areas, simultaneously. Notably, by comparing Fig. \ref{Fig5}a
and Fig. \ref{Fig5}b, together, we can see that the electrical charge acts
the opposite of ModMax's parameter.

iv) Our findings in Fig. \ref{Fig5}c, indicate that $S_{root}$, and $%
S_{div_{1}}$ are not very sensitive to changes of $f_{R_{0}}$. However, $%
S_{div_{2}}$ increases by increasing $f_{R_{0}}$. It means that the local
stability area of large black holes in $F(R)$-ModMax theory decreases.

v) $S_{root}$ and $S_{div_{1}}$ change by varying $\left\vert
R_{0}\right\vert $, but these changes are less than $S_{div_{2}}$. In other
words, by increasing $\left\vert R_{0}\right\vert $, the second divergence
point ($S_{div_{2}}$) decreases, which leads to increasing the local
stability area.

\subsection{Helmholtz Free Energy}

The global stability of a thermodynamic system is determined by the negative
value of the Helmholtz free energy in the canonical ensemble. In order to
examine the global stability of black holes in $F(R)$-ModMax theory, our
objective is to assess the Helmholtz free energy. It is worth mentioning
that, in the usual thermodynamics, the Helmholtz free energy is typically
defined as 
\begin{equation}
F=U-TS,
\end{equation}%
that in relation to the black holes ($U=M$), it turns to the following
relation 
\begin{equation}
F(T,Q)=M\left( S,Q\right) -TS.
\end{equation}%
Using Eqs. (\ref{MSQ}) and (\ref{TM}), we can get the Helmholtz free energy,
which yields 
\begin{equation}
F(T,Q)=\frac{\left( S+3\pi Q^{2}e^{-\gamma }\right) \left(
1+f_{R_{0}}\right) +\frac{R_{0}S^{2}}{12\pi }}{4\sqrt{\pi S\left(
1+f_{R_{0}}\right) }}.  \label{F}
\end{equation}

To study the global stability of black holes, we have to determine the
negative value of the Helmholtz free energy. To achieve this objective, we
can determine the roots of the Helmholtz free energy (\ref{F}) by solving
the equation $F(T,Q)=0$, which leads to 
\begin{equation}
S_{F=0}=\frac{-6\pi \left( 1+f_{R_{0}}\right) }{R_{0}}\left[ 1+\sqrt{1-\frac{%
R_{0}Q^{2}e^{-\gamma }}{1+f_{R_{0}}}}\right] ,
\end{equation}%
which indicates that there is only one real root of the Helmholtz free
energy if we respect the constraint $R_{0}\leq \frac{1+f_{R_{0}}}{%
Q^{2}e^{-\gamma}}$. Notably, for $R_{0}<0$, this constraint is automatically
satisfied.

\begin{figure}[tbph]
\centering
\includegraphics[width=0.35\linewidth]{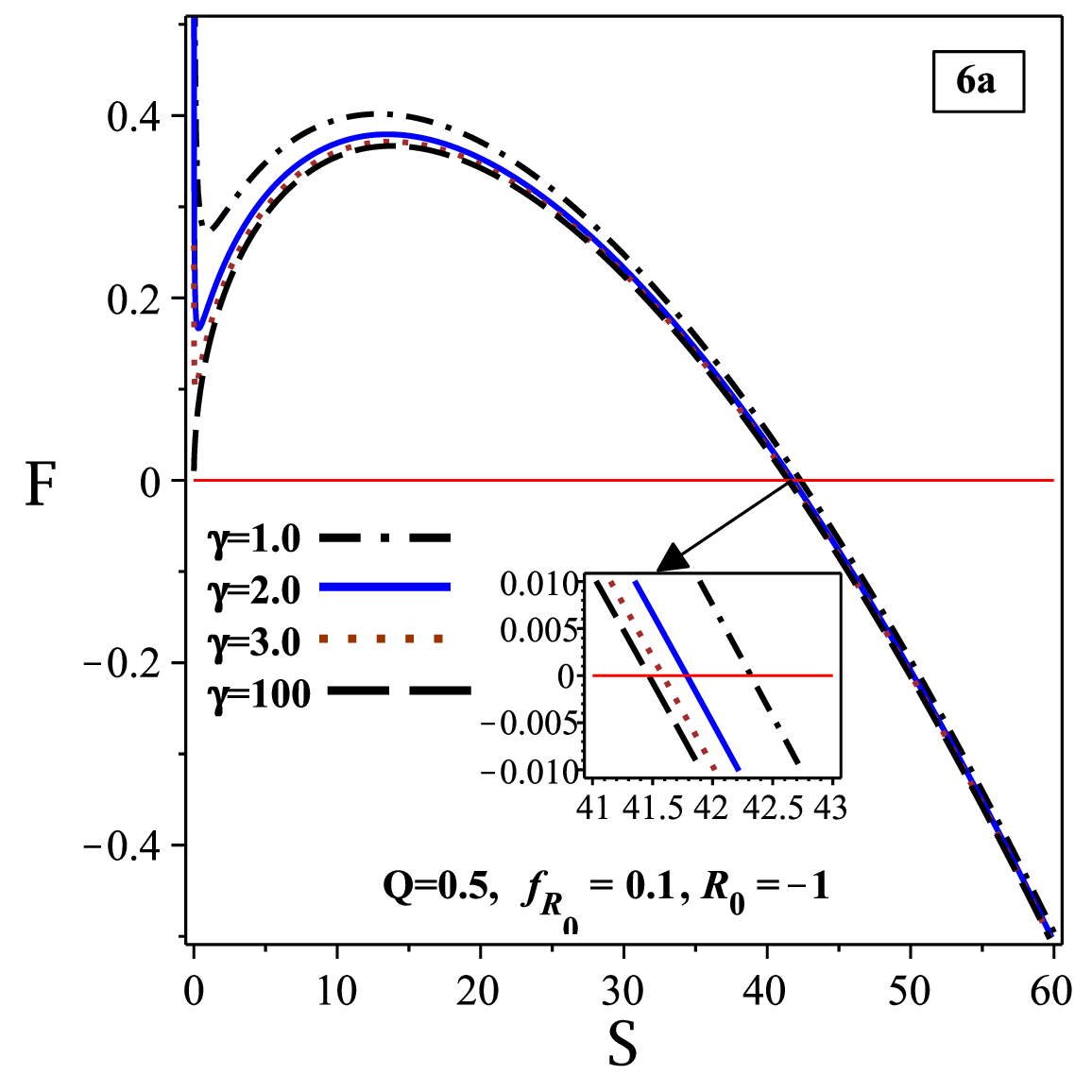} \includegraphics[width=0.35%
\linewidth]{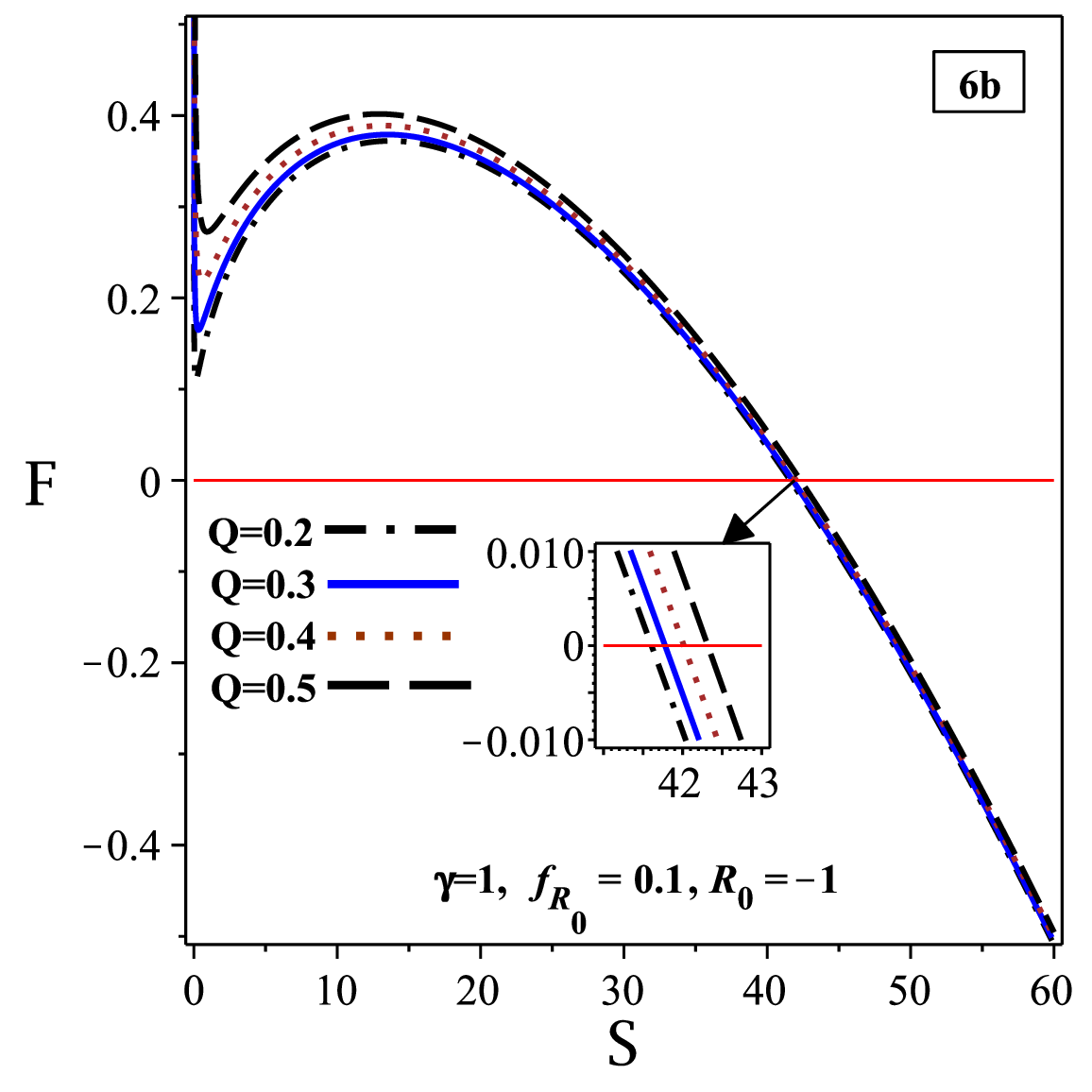}\newline
\includegraphics[width=0.35\linewidth]{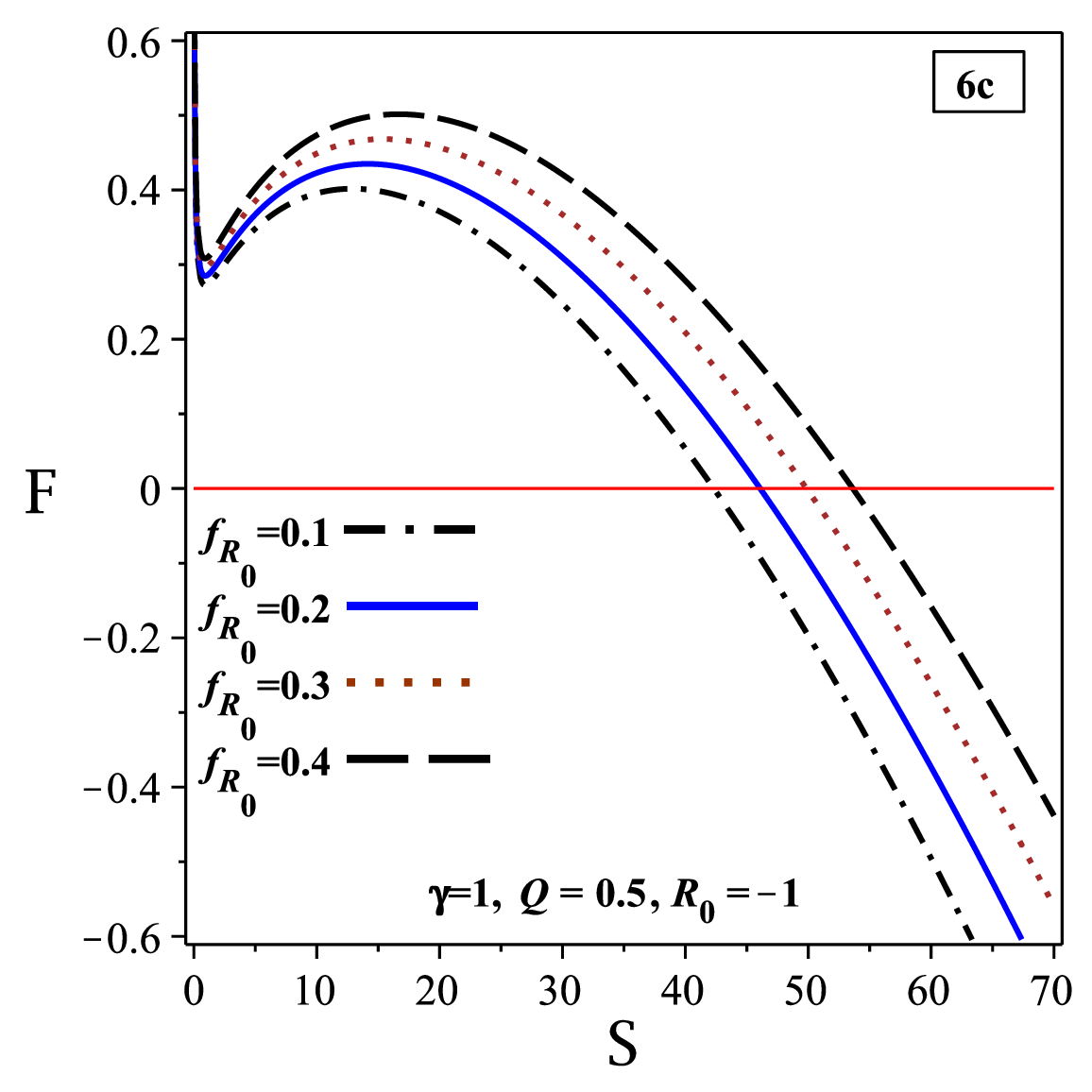} \includegraphics[width=0.35%
\linewidth]{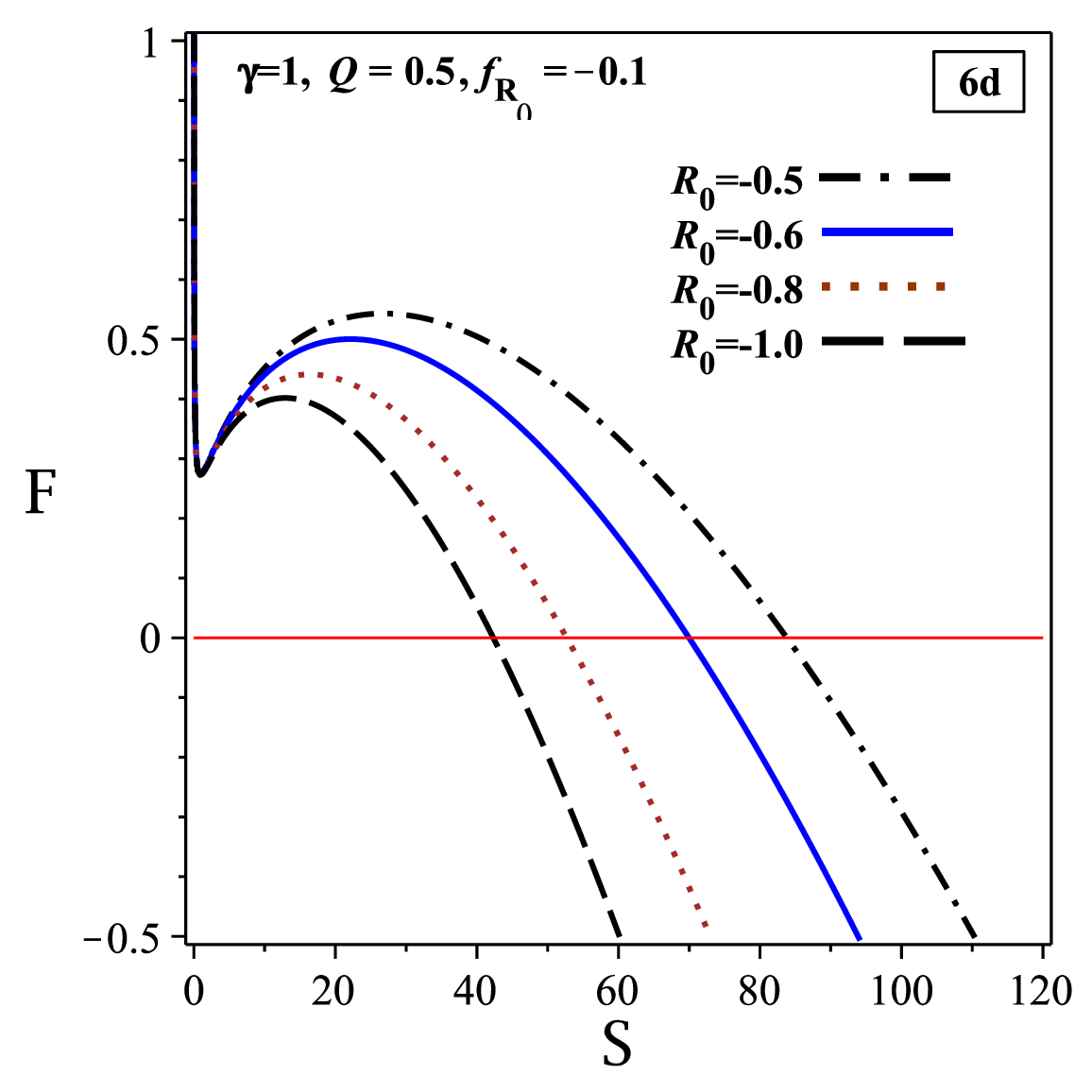}\newline
\caption{The Helmholtz free energy $F$ versus $S$ for different values of
the parameters.}
\label{Fig6}
\end{figure}

In order to observe the impact of different parameters on the global
stability regions of black holes, we graph the Helmholtz free energy against 
$S$ in four sections of Figure \ref{Fig6}. Our results are:

i) The global stability area is located at $S>S_{F=0}$. Indeed, there are
two different areas, before and after of the root of Helmholtz free energy ($%
S<S_{F=0}$, and $S>S_{F=0}$). The Helmholtz free energy is positive in the
range $S<S_{F=0}$, which indicates that the black holes cannot satisfy the
global stability. In other words, the small black hole in $F(R)$-ModMax
theory is not a global stable system. Conversely, the Helmholtz free energy
exhibits a negative value within the range of $S>S_{F=0}$. So, the large
black holes are global stable systems (see all panels in Fig. \ref{Fig6}).

ii) Our analysis of Fig. \ref{Fig6}a and Fig. \ref{Fig6}d, indicate that by
increasing $\gamma $ and $\left\vert R_{0}\right\vert $, the root of the
Helmholtz free energy ($S_{F=0}$) decreases, which leads to increasing the
global stability area.

iii) By increasing $Q$ and $f_{R_{0}}$, the global stability area decreases
because $S_{F=0}$ increases (see Fig. \ref{Fig6}b and Fig. \ref{Fig6}c, for
more details).

As a result, we found that the large black holes in $F(R)$-ModMax theory can
satisfy the local and global stability conditions, simultaneously. This
result is extracted by comparing the areas of the local and global
stabilities in Figs. \ref{Fig5}, and \ref{Fig6}.

\section{Geometrical Thermodynamics}

Geometrical thermodynamics (GTD) offers an alternative method to investigate
the critical points of phase transition in black holes. GTD's approach uses
the thermodynamic quantities to build a metric that describes the thermal
phases of black hole \cite%
{WeinholdI,WeinholdII,RuppeinerI,RuppeinerII,RuppeinerIII,QuevedoI,QuevedoII,GTD1,GTD2,GTD3,GTD4,HPEM,HPEM1,HPEM2,HPEM3,Rodrigues2012,Han2012,Azreg2014,Suresh2015,Mo2016,Quevedo2016,Channuie2018,Bhattacharya2019,HPEM4,Mansoori2019,Upadhyay2021,HPEM5}%
. The thermodynamic behavior in the GTD method is determined by analyzing
the Ricci scalar of the thermodynamic metric. Specifically, the changes in
the sign of the Ricci scalar and its divergences are utilized to represent
the different thermal phases of black holes. In essence, the objective is to
characterize the thermodynamic properties of black holes using Riemannian
calculus. In this regard, some methods are introduced for constructing the
thermodynamical metric, for example, Weinhold \cite{WeinholdI,WeinholdII},
Ruppeiner \cite{RuppeinerI,RuppeinerII}, Quevedo \cite{QuevedoI,QuevedoII},
Fisher-Rau \cite{QuevedoI}, and HPEM \cite{HPEM} are some of these
thermodynamical metrics. GTD of black holes has been evaluated with various
thermodynamic criteria, each of which has been associated with success and
failure (see Refs. \cite{HPEM4,HPEM5,Mahmoudi2023}, for more details).

Previous research has examined the limitations of Ruppeiner, Weinhold,
Fisher-Rau, and Quevedo metrics in accurately assessing the thermal phases
of certain types of black holes (see Refs. \cite{HPEM1,HPEM2,HPEM4} for
further information). Consequently, we employ the HPEM's metric to explore
the thermal phases of electrically charged black holes in the $F(R)$-ModMax
theory.

The HPEM's metric is introduced as \cite{HPEM} 
\begin{equation}
dS_{HPEM}^{2}=\frac{SM_{S}}{M_{QQ}^{3}}\left(
-M_{SS}dS^{2}+M_{QQ}dQ^{2}\right) ,  \label{HPEM}
\end{equation}%
where $M_{S}=$ $\left( \frac{\partial M(S,Q)}{\partial S}\right) _{Q}$, $%
M_{SS}=\left( \frac{\partial ^{2}M(S,Q)}{\partial S^{2}}\right) _{Q}$ and $%
M_{QQ}=\left( \frac{\partial ^{2}M(S,Q)}{\partial Q^{2}}\right) _{S}$. After
some calculation, we can find the numerator and denominator of the Ricci
scalar of HPEM's metric in the following forms 
\begin{eqnarray}
\text{numerator }(R_{HPEM}) &=&S^{2}M_{S}^{2}M_{QQQ}^{3}M_{SSS}\left( \frac{%
M_{SS}}{M_{S}}-\frac{1}{S}\right) +S^{2}M_{S}^{2}M_{QQ}^{3}M_{SQQ}\left(
2M_{SSS}+\frac{M_{SS}}{S}-\frac{M_{SS}^{2}}{M_{S}}\right)  \notag \\
&&+S^{2}M_{S}^{2}M_{SS}^{2}M_{QQQ}^{2}\left( \frac{M_{SQ}M_{QQ}}{M_{S}M_{QQQ}%
}-9\right) +6S^{2}M_{S}^{2}M_{QQ}^{2}M_{SS}^{2}\left( \frac{M_{QQQQ}}{M_{QQ}}%
-\frac{M_{SSQQ}}{M_{SS}}\right)  \notag \\
&&+S^{2}M_{SQ}^{2}M_{SS}^{2}M_{QQ}^{2}\left( 2-\frac{M_{S}M_{SSQ}}{%
M_{SS}M_{SQ}}\right) +S^{2}M_{QQ}^{2}\left(
M_{S}^{2}M_{SSQ}^{2}-2M_{SS}^{3}M_{QQ}\right)  \notag \\
&&+S^{2}M_{S}^{2}M_{SS}M_{QQ}\left( 2M_{SQQ}^{2}+4M_{QQQ}M_{SSQ}\right)
-2M_{S}^{2}M_{SS}M_{QQ}^{3},  \label{num} \\
&&  \notag \\
\text{denominator }(R_{HPEM}) &=&2S^{3}M_{S}^{3}M_{SS}^{2},  \label{dnom}
\end{eqnarray}%
in which $M_{XX}=\left( \frac{\partial ^{2}M}{\partial X^{2}}\right) $, $%
M_{XY}=\left( \frac{\partial ^{2}M}{\partial X\partial Y}\right) $, $%
M_{XXX}=\left( \frac{\partial ^{3}M}{\partial X^{3}}\right) $, $%
M_{XXXX}=\left( \frac{\partial ^{4}M}{\partial X^{4}}\right) $, and $%
M_{XXYY}=\left( \frac{\partial ^{4}M}{\partial X^{2}\partial Y^{2}}\right) $.

\begin{figure}[tbph]
\centering
\includegraphics[width=0.35\linewidth]{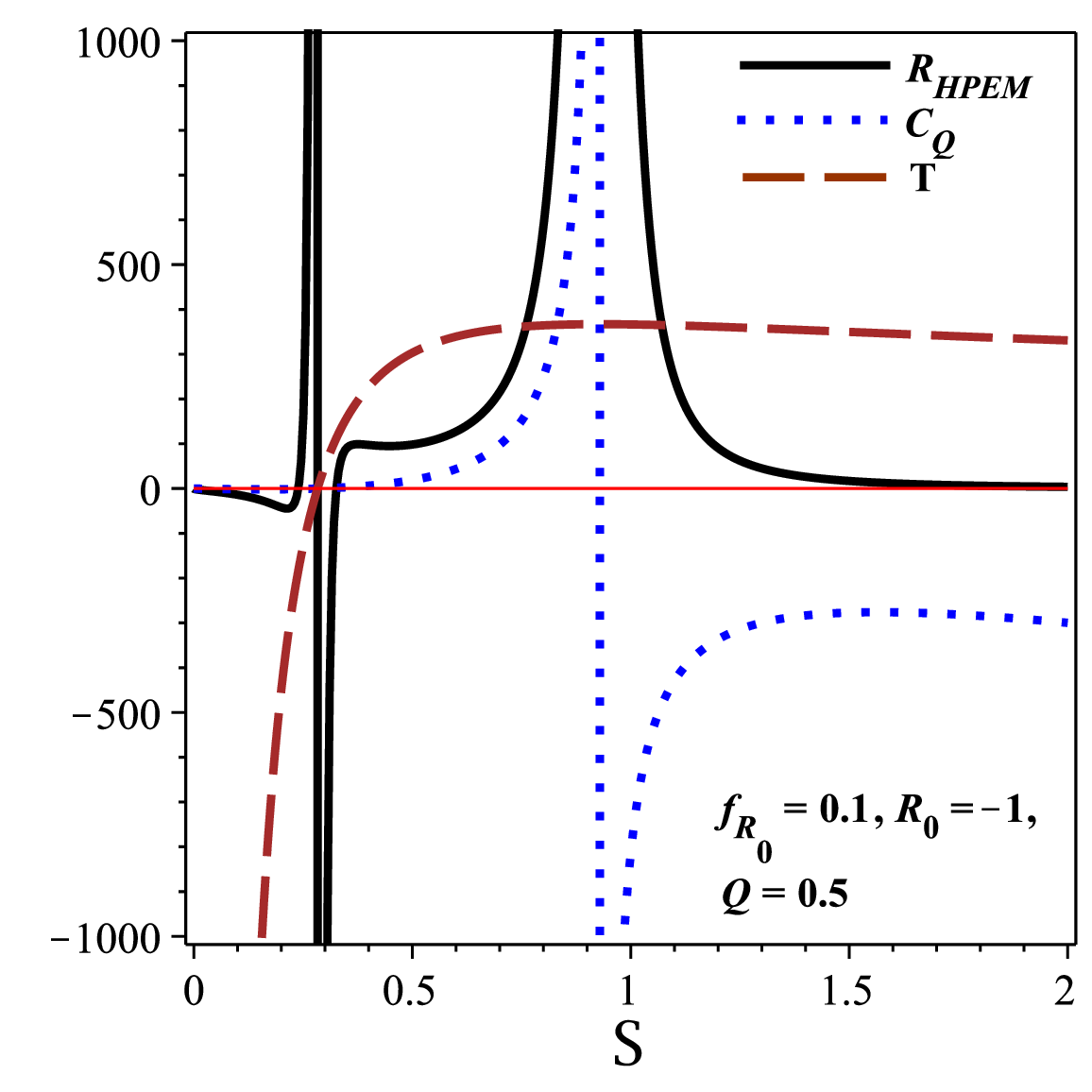} \includegraphics[width=0.35%
\linewidth]{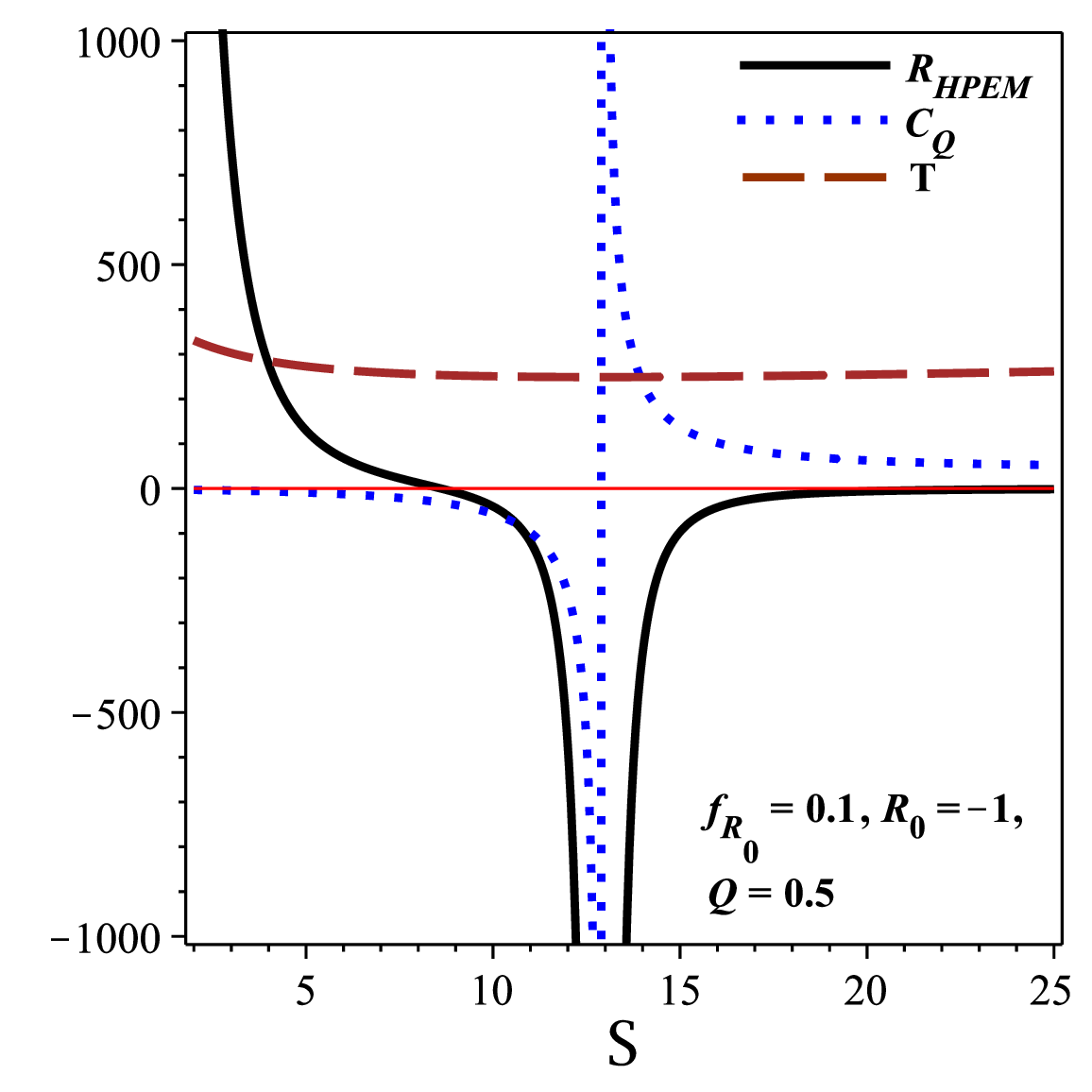}\newline
\caption{Ricci scalar of HPEM's metric $R_{HPEM}$, the heat capacity $C_{Q}$%
, and the temperature $T$ versus $S$ for different values of the parameters.}
\label{Fig7}
\end{figure}

Our findings, in Fig. \ref{Fig7}, reveal that the divergence points of the
Ricci scalar of HPEM's metric coincides completely with both the phase
transition critical and the physical limitation points of the heat capacity.
So, all the thermodynamic criteria are included in the divergences of the
Ricci scalar of HPEM's metric. In addition, the divergences of the Ricci
scalar of HPEM's metric are different before and after the physical
limitation points with the phase transition critical points. Indeed. The
sign of the Ricci scalar of HPEM's metric changes before and after
divergence, which is related to the physical limitation point. Nevertheless,
the signs of the Ricci scalar exhibit identical characteristics in the
vicinity of the critical points of the phase transition. These divergences
are known as $\Lambda $ divergences. So, by adopting this methodology, we
can differentiate between physical constraints and critical points of phase
transitions.

\section{Conclusions}

In this work, we have exacted analytical solutions in $F(R)$-ModMax theory
of gravity. Then, we calculated the Kretschmann scalar of the obtained
analytical solutions in order to find an essential singularity. Our results
indicated there was a curvature singularity at $r=0$. In order to determine
the roots, a graphical representation of the metric function has been
plotted against the variable $r$ in Figure \ref{Fig1}. Our findings revealed
that for $R_{0}>0$, there were three different roots, an inner root, an
event horizon, and a cosmological horizon. If $R_{0}<0$ and certain
parameters are adjusted, the solution could exhibit an inner horizon and an
event horizon (two roots), a single root (in the extreme case), or a naked
singularity. Our analysis indicated that we could have an event horizon that
covered the singularity at $r=0$ by adjusting the parameters. So, the
obtained solution could be related to the black hole solution within the $%
F(R)$-ModMax theory.

The thermodynamic quantities of charged black holes within the $F(R)$-ModMax
theory were calculated, and the validity of the first law of thermodynamics
was checked. Our analysis indicated that the obtained conserved and
thermodynamic quantities of these black holes (such as the Hawking
temperature, the electrical charge, the electrical potential, entropy, and
the total mass) satisfied the first law of thermodynamics. Furthermore, an
assessment was conducted to analyze the impact of different factors on both
the Hawking temperature and the total mass. Our results revealed there was
one real root for the Hawking temperature, which is dependent on different
parameters. It is notable that, this root was removed in the limit $\gamma
\rightarrow \infty $. By considering $R_{0}<0$ and different values of
parameters, the temperature was positive for large black holes. In addition,
there were two extremum points for the Hawking temperature (except in the
limit $\gamma \rightarrow \infty $). The total mass was always positive and
there was one extremum point that depended on different parameters. Notably,
there was no extremum point for the total mass when $\gamma \rightarrow
\infty $.

We have studied the heat capacity of the charged black holes in $F(R)$%
-ModMax theory to investigate the local stability. We found two critical
points (or phase transition points), and one physical limitation point for
the heat capacity. These points depended on the parameters of the system.
Our findings indicated that there were two local stability and physical
areas, simultaneously. These areas are located at $S_{root}<S<S_{div_{1}}$,
and $S>S_{div_{2}}$. In other words, the electrical charged black holes in $%
F(R)$-ModMax theory of gravity could be stable and physical objects,
simultaneously, when are located in the ranges $S_{root}<S<S_{div_{1}}$ (or
medium black holes) and $S>S_{div_{2}}$ (or large black holes).

We have examined the Helmholtz free energy ($F$) to assess the global
stability of the acquired black holes. We found that there was one real root
of the Helmholtz free energy ($S_{F=0}$), which depended on the parameters
of $F(R)$ gravity (i.e., $R_{0}$, and $f_{R_{0}}$), the electrical charge $Q$%
, and ModMax's parameter ($\gamma$). The global (in)stability area was
located at $S>S_{F=0}$ ($S<S_{F=0}$). In other words, there were two
different areas, before and after the root of Helmholtz free energy ($%
S<S_{F=0}$, and $S>S_{F=0}$). The Helmholtz free energy was positive in the
range $S<S_{F=0}$, which indicated that the black holes could not satisfy
the global stability. Conversely, the Helmholtz free energy exhibited a
negative value within the range of $S>S_{F=0}$. Therefore, the large black
holes were global stable systems.

In the last section, we have studied geometrical thermodynamics by using
HPEM's metric for the obtained electrical charge black holes in $F(R)$%
-ModMax theory. In Fig. \ref{Fig7}, it was observed that the Ricci scalar of
HPEM's metric exhibited divergence points that aligned with both phase
transition critical and the physical limitation points of the heat capacity.
Furthermore, the divergences of the Ricci scalar before and after the
physical limitation points were distinct from those at the phase transition
critical points. Indeed, the alteration in the Ricci scalar sign of HPEM's
metric occurred before and after divergence, indicating a connection to the
physical limitation point. Nevertheless, the Ricci scalar signs remained
consistent around the critical points of phase transition. Consequently, the
utilization of HPEM's metric allowed us to differentiate between phase
transition critical and physical limitation points.

Given the importance of the NED field in various aspects of physics,
the study of modified gravitational theories such as the $F(R)$
theory with NED fields or $F(R)$ models inspired by NED (which can
describe some phenomena observed in astrophysics and cosmology) is very
interesting. For example, Born-Infeld-$F(R)$ gravity \cite%
{Makarenko2014,Makarenkoa2014} has shown some interesting properties in
cosmology and black holes. Born-Infeld-$F(R)$ gravity opened new
ways to answer some questions of gravitational dynamics at low energies.
This theory efficiently explains the structure of stars without the need to
reconsider the convenient approximation of a perfect fluid. We can therefore
extend our model by a Lagrangian reform (similar to the Born-Infeld-$F(R)$ theory) and explore cosmological and astrophysical applications in
the future.

\begin{acknowledgements}
	B. Eslam Panah thanks University of Mazandaran.
\end{acknowledgements}


\begin{thebibliography}{999}
\bibitem{NojiriO2003} S. Nojiri, and S. D. Odintsov, Phys. Rev. D \textbf{68}%
, 123512 (2003).

\bibitem{NojiriO2011} S. Nojiri, and S. D. Odintsov, Phys. Rept. \textbf{505}%
, 59 (2011).

\bibitem{F(R)I} H. A. Buchdahl, Mon. Not. Roy. Astron. Soc. \textbf{150}, 1
(1970).

\bibitem{F(R)II} S. M. Carroll, V. Duvvuri, M. Trodden, and M. S. Turner,
Phys. Rev. D \textbf{70}, 043528 (2004).

\bibitem{F(R)IIb} S. Capozziello, S. Nojiri, and S. D. Odintsov, Phys. Lett.
B \textbf{634}, 93 (2006).

\bibitem{F(R)III} T. P. Sotiriou, and V. Faraoni, Rev. Mod. Phys. \textbf{82}%
, 451 (2010).

\bibitem{F(R)IV} S. Nojiri, S. D. Odintsov, and V. K. Oikonomou, Phys. Rept. 
\textbf{692}, 1 (2017).

\bibitem{Mod1} A. A. Starobinsky, Phys. Lett. B \textbf{91}, 99 (1980).

\bibitem{Mod2} I. Sawicki, and W. Hu, Phys. Rev. D \textbf{75}, 127502
(2007).

\bibitem{Mod3} L. Amendola, and S. Tsujikawa, Phys. Lett. B \textbf{660},
125 (2008).

\bibitem{Mod4} S. Tsujikawa, Phys. Rev. D \textbf{77}, 023507 (2008).

\bibitem{Mod5} G. Cognola, E. Elizalde, S. Nojiri, S. D. Odintsov, L.
Sebastiani, and S. Zerbini, Phys. Rev. D \textbf{77}, 046009 (2008).

\bibitem{Mod6} S. Capozziello, E. Piedipalumbo, C. Rubano, and P.
Scudellaro, Astron. Astrophys. \textbf{505}, 21 (2009).

\bibitem{Mod7} A. V. Astashenok, S. Capozziello, and S. D. Odintsov, JCAP 
\textbf{12}, 040 (2013).

\bibitem{Mod7a} S. D. Odintsov, and V. K. Oikonomou, Phys. Lett. B \textbf{%
833}, 137353 (2022).

\bibitem{Mod7b} S. D. Odintsov, V. K. Oikonomou, and G. S. Sharov, Phys.
Lett. B \textbf{843}, 137988 (2023).

\bibitem{Mod8} M. Leizerovich, L. Kraiselburd, S. J. Landau, and C. G.
Scoccola, Phys. Rev. D \textbf{105}, 103526 (2022).

\bibitem{Capozziello2002} S. Capozziello, Int. J. Mod. Phys. D \textbf{11},
483 (2002).

\bibitem{Carroll2004} S. M. Carroll, V. Duvvuri, M. Trodden, and M. S.
Turner, Phys. Rev. D \textbf{70}, 043528 (2004).

\bibitem{Sotiriou2006} T. P. Sotiriou, Class. Quantum Grav. \textbf{23},
5117 (2006).

\bibitem{Hu2007} W. Hu, and I. Sawicki, Phys. Rev. D \textbf{76}, 064004
(2007).

\bibitem{Baghram2007} S. Baghram, M. Farhang, and S. Rahvar, Phys. Rev. D 
\textbf{75}, 044024 (2007).

\bibitem{NojiriO2008} S. Nojiri, and S. D. Odintsov, Phys. Rev. D \textbf{77}
,026007 (2008).

\bibitem{Cognola2008} G. Cognola, E. Elizalde, S. Nojiri, S. D. Odintsov, L.
Sebastiani, and S. Zerbini, Phys. Rev. D \textbf{77}, 046009 (2008).

\bibitem{Elizalde2010} E. Elizalde, S. Nojiri, S. D. Odintsov, and D.
Saez-Gomez, Eur. Phys. J. C \textbf{70}, 351 (2010).

\bibitem{NojiriOdark2006} S. Nojiri, and S. D. Odintsov, Phys. Rev. D 
\textbf{74}, 086005 (2006).

\bibitem{CapozzielloCT2007} S. Capozziello, V. F. Cardone, and A. Troisi,
Mon. Not. Roy. Astron. Soc. \textbf{375}, 1423 (2007).

\bibitem{CapozzielloH2013} S. Capozziello, T. Harko, T. S. Koivisto, F. S.
N. Lobo, and G. J. Olmo, JCAP \textbf{07}, 024 (2013).

\bibitem{CapozzielloI} S. Capozziello, and A. Troisi, Phys. Rev. D \textbf{72%
}, 044022 (2005).

\bibitem{CapozzielloII} S. Capozziello, A. Stabile, and A. Troisi, Phys.
Rev. D \textbf{76}, 104019 (2007).

\bibitem{NED1} W. Heisenberg, and H. Euler, Z. Phys. \textbf{98}, 714 (1936).

\bibitem{NED2} J. Schwinger, Phys. Rev. \textbf{82}, 664 (1951).

\bibitem{NED3} H. Yajima, and T. Tamaki, Phys. Rev. D \textbf{63}, 064007
(2001).

\bibitem{NEDMagI} A. Ibrahim, et al., Astrophys. J. Lett. \textbf{574}, L51
(2002).

\bibitem{NEDMagII} H. J. Mosquera Cuesta, and J. M. Salim, Mon. Not. R.
Acad. Sci. \textbf{354}, L55 (2004).

\bibitem{Bardeen1968} J. M. Bardeen, Proceedings of International Conference
GR5, (USSR, Tbilisi, Georgia), 174 (1968).

\bibitem{Ayon1998} E. Ayon-Beato, and A. Garcia, Phys. Rev. Lett. \textbf{80}%
, 5056 (1998).

\bibitem{Chinaglia2017} S. Chinaglia, and S. Zerbini, Gen. Rel. Grav. 
\textbf{49}, 75 (2017).

\bibitem{Nojiri2017} S. Nojiri, and S.\thinspace D. Odintsov, Phys. Rev. D 
\textbf{96}, 104008 (2017).

\bibitem{BigBang1} E. Ayon-Beato, and A. Garcia, Phys. Lett. B \textbf{464},
25 (1999).

\bibitem{BigBang2} I. Dymnikova, Class. Quantum Grav. \textbf{21}, 4417
(2004).

\bibitem{BigBang3} C. Corda, and H. J. Mosquera Cuesta, Mod. Phys. Lett. A 
\textbf{25}, 2423 (2010).

\bibitem{NEDI} V. A. De Lorenci, and M. A. Souza, Phys. Lett. B \textbf{512}%
, 417 (2001).

\bibitem{NEDII} V. A. De Lorenci, and R. Klippert, Phys. Rev. D \textbf{65},
064027 (2002).

\bibitem{NEDIII} M. Novello, et al., Class. Quantum Grav. \textbf{20}, 959
(2003).

\bibitem{NEDIV} M. Novello, and E. Bittencourt, Phys. Rev. D \textbf{86},
124024 (2012).

\bibitem{NEDP1} Z. Bialynicka-Birula, and I. Bialynicka-Birula, Phys. Rev. D 
\textbf{2}, 2341 (1970).

\bibitem{NEDP2} H. J. Mosquera, and J. M. Cuesta Salim, Astrophys. J. 
\textbf{608}, 925 (2004).

\bibitem{BI} M. Born, and L. Infeld, Proc. Royal Soc. (London) A \textbf{144}%
, 425 (1934).

\bibitem{PM1} M. Hassaine, and C. Martinez, Phys. Rev. D \textbf{75}, 027502
(2007).

\bibitem{PM2} H. Maeda, M. Hassaine, and C. Martinez, Phys. Rev. D \textbf{79%
}, 044012 (2009).

\bibitem{PM3} S. H. Hendi, Phys. Lett. B \textbf{678}, 438 (2009).

\bibitem{PM5} M. M. Stetsko, Phys. Rev. D \textbf{99}, 044028 (2019).

\bibitem{RemovePM1} B. Eslam Panah, EPL \textbf{134}, 20005 (2021).

\bibitem{RemovePM2} S. H. Mazharimousavi, Mod. Phys. Lett. A \textbf{37},
2250170 (2022).

\bibitem{ModMaxI} I. Bandos, K. Lechner, D. Sorokin, and P. K. Townsend,
Phys. Rev. D \textbf{102}, 121703 (2020).

\bibitem{BH1} T. Multamaki, and I. Vilja, Phys. Rev. D \textbf{74}, 064022
(2006).

\bibitem{BH2} A. de la Cruz-Dombriz, A. Dobado, and A. L. Maroto, Phys. Rev.
D \textbf{80}, 124011 (2009).

\bibitem{BH3} S. Capozziello, M. De laurentis, and A. Stabile, Class.
Quantum Grav. \textbf{27}, 165008 (2010).

\bibitem{BH4} W. Nelson, Phys. Rev. D \textbf{82}, 104026 (2010).

\bibitem{BH5} T. Moon, Y. S. Myung, and E. J. Son, Gen. Relativ. Gravit. 
\textbf{43}, 3079 (2011).

\bibitem{BH6} L. Sebastiani, and S. Zerbini, Eur. Phys. J. C \textbf{71},
1591 (2011).

\bibitem{BH7} S. H. Mazharimousavi, and M. Halilsoy, Phys. Rev. D \textbf{84}%
, 064032 (2011).

\bibitem{BH8} S. H. Hendi, B. Eslam Panah, and S. M. Mousavi, Gen. Relativ.
Gravit. \textbf{44}, 835 (2012).

\bibitem{BH9} D. Bazeia, L. Losano, Gonzalo J. Olmo, and D. Rubiera-Garcia,
Phys. Rev. D \textbf{90}, 044011 (2014).

\bibitem{BH10} S. H. Hendi, B. Eslam Panah, and R. Saffari, Int. J. Mod.
Phys. D \textbf{23}, 1450088 (2014).

\bibitem{BH11} A. Kehagias, C. Kounnas, D. Lust, and A. Riotto, JHEP \textbf{%
05}, 143 (2015).

\bibitem{BH12} M. E. Rodrigues, E. L. B. Junior, G. T. Marques, and V. T.
Zanchin, Phys. Rev. D \textbf{94}, 024062 (2016).

\bibitem{BH13} P. Canate, L. G. Jaime, and M. Salgado, Class. Quantum Grav. 
\textbf{33}, 155005 (2016).

\bibitem{BH14} A. K. Mishra, M. Rahman, and S. Sarkar, Class. Quantum Grav. 
\textbf{35}, 145011 (2018).

\bibitem{BH15} J. Sultana, and D. Kazanas, Gen. Relativ. Gravit. \textbf{50}%
, 137 (2018).

\bibitem{BH16} G. G. L. Nashed, and S. Capozziello, Phys. Rev. D \textbf{99}%
, 104018 (2019).

\bibitem{BH17} M. Zhang, and R. B. Mann, Phys. Rev. D \textbf{100}, 084061
(2019).

\bibitem{BH18} G. G. L. Nashed, and E. N. Saridakis, Phys. Rev. D \textbf{102%
}, 124072 (2020).

\bibitem{BH19} G. G. L. Nashed, and K. Bamba, PTEP \textbf{2020}, 043E05
(2020).

\bibitem{BH20} E. Elizalde, et al., Eur. Phys. J. C \textbf{80}, 109 (2020).

\bibitem{BH21} G. G. L. Nashed, W. El Hanafy, S. D. Odintsov, and V. K.
Oikonomou, Int. J. Mod. Phys. D \textbf{29}, 2050090 (2020).

\bibitem{BH22} S. C. Jaryal, and A. Chatterjee, Eur. Phys. J. C \textbf{81},
273 (2021).

\bibitem{BH23} E. F. Eiroa, and G. Figueroa-Aguirre, Phys. Rev. D \textbf{103%
}, 044011 (2021).

\bibitem{BH24} Z. Y. Tang, B. Wang, T. Karakasis, and E. Papantonopoulos,
Phys. Rev. D \textbf{104}, 064017 (2021).

\bibitem{BH25} T. Karakasis, E. Papantonopoulos, Z. Y. Tang, and B. Wang,
Eur. Phys. J. C \textbf{81}, 897 (2021).

\bibitem{BH26} B. Eslam Panah, J. Math. Phys. \textbf{63}, 112502 (2022).

\bibitem{BH27} M. E. Rodrigues, E. L. B. Junior, G. T. Marques, and Julio C.
Fabris, Eur. Phys. J. C \textbf{76}, 250 (2016).

\bibitem{BH28} S. Kanzi, I. Sakall\i , and B. Pourhassan, Symmetry. \textbf{%
15}, 873 (2023).

\bibitem{Bekenstein1973} J. D. Bekenstein, Phys. Rev. D \textbf{7}, 2333
(1973).

\bibitem{Hawking1974} S. W. Hawking, Nature (London) \textbf{248}, 30 (1974).

\bibitem{Bardeen1973} J. M. Bardeen, B. Carter, and S.W. Hawking, Comm.
Math. Phys. \textbf{31}, 161 (1973).

\bibitem{The1} Y. S. Myung, Y. -W. Kim, and Y. -J. Park, Phys. Rev. D 
\textbf{78}, 084002 (2008).

\bibitem{The2} R. Banerjee, S. Ghosh, and D. Roychowdhury, Phys. Lett. B 
\textbf{696}, 156 (2011).

\bibitem{The3} B. P. Dolan, D. Kastor, D. Kubiznak, R. B. Mann, and J.
Traschen, Phys. Rev. D \textbf{87}, 104017 (2013).

\bibitem{The4} N. Altamirano, D. Kubiznak, and R. B. Mann, Phys. Rev. D 
\textbf{88}, 101502 (2013).

\bibitem{The5} L. -C. Zhang, M. -S. Ma, H. -H. Zhao, and R. Zhao, Eur. Phys.
J. C \textbf{74}, 3052 (2014).

\bibitem{The6} J. -L. Zhang, R. -G. Cai, and H. Yu, Phys. Rev. D \textbf{91}%
, 044028 (2015).

\bibitem{The7} J. -X. Mo, G. -Q. Li, and X. -B. Xu, Eur. Phys. J. C \textbf{%
76}, 545 (2016).

\bibitem{The8} S. H. Hendi, S. Panahiyan, B. Eslam Panah, and Z. Armanfard,
Eur. Phys. J. C \textbf{76}, 396 (2016).

\bibitem{The9} G. -Q. Li, and J. -X. Mo, Phys. Rev. D \textbf{93}, 124021
(2016).

\bibitem{The10} D. Hansen, D. Kubiznak, and R. B. Mann, JHEP \textbf{01},
047 (2017).

\bibitem{The11} R. A. Hennigar, D. Kubiznak, and R. B. Mann, Phys. Rev. D 
\textbf{100}, 064055 (2019).

\bibitem{The12} S. -W. Wei, Y. -X. Liu, and Y. -Q. Wang, Phys. Rev. D 
\textbf{99}, 044013 (2019).

\bibitem{The13} P. Wang, H. Wu, and H. Yang, JCAP \textbf{04}, 052 (2019).

\bibitem{The14} A. Anabalon, F. Gray, R. Gregory, D. Kubiznak, and R. B.
Mann, JHEP \textbf{04}, 096 (2019).

\bibitem{The15} M. Cataldo, P. A. Gonz\'{a}lez, J. Saavedra, Y. V\'{a}zquez,
and B. Wang, Phys. Rev. D \textbf{103}, 024047 (2021).

\bibitem{The16} M. Tataryn, and M. Stetsko, Gen. Relativ. Gravit. \textbf{53}%
, 72 (2021).

\bibitem{The17} S. -W. Wei, and Y. -X. Liu, Phys. Rev. D \textbf{105},
104003 (2022).

\bibitem{The18} A. Sood, A. Kumar, J. K. Singh, and S. G. Ghosh, Eur. Phys.
J. C \textbf{82}, 227 (2022).

\bibitem{The19} E. Hirunsirisawat, R. Nakarachinda, and C. Promsiri, Phys.
Rev. D \textbf{105}, 124049 (2022).

\bibitem{The20} N. -C. Bai, L. Li, and J. Tao, Phys. Rev. D \textbf{107},
064015 (2023).

\bibitem{The21} A. M. Frassino, J. F. Pedraza, A. Svesko, and M. R. Visser,
Phys. Rev. Lett. \textbf{130}, 161501 (2023).

\bibitem{The22} Z. -F. Mai, R. Xu, D. Liang, and L. Shao, Phys. Rev. D 
\textbf{108}, 024004 (2023).

\bibitem{The23} B. Hamil, and B. C. Lutfuoglu, Phys. Dark Universe. \textbf{%
42}, 101293 (2023).

\bibitem{The24} D. Chen, Y. He, and J. Tao, Eur. Phys. J. C \textbf{83}, 872
(2023).

\bibitem{HeatC1} B. P. Dolan, Class. Quantum Gravit. \textbf{31}, 165011
(2014).

\bibitem{HeatC2} S. Grunau, and H. Neumann, Class. Quantum Gravit. \textbf{32%
}, 175004 (2015).

\bibitem{HeatC3} B. Eslam Panah, S. H. Hendi, S. Panahiyan, and M. Hassaine,
Phys. Rev. D \textbf{98}, 084006 (2018).

\bibitem{WeinholdI} F. Weinhold, J. Chem. Phys. \textbf{63}, 2479 (1075).

\bibitem{WeinholdII} F. Weinhold, J. Chem. Phys. \textbf{63}, 2484 (1975).

\bibitem{RuppeinerI} G. Ruppeiner, Phys. Rev. A \textbf{20}, 1608 (1979).

\bibitem{RuppeinerII} G. Ruppeiner, Rev. Mod. Phys. \textbf{67}, 605 (1995).

\bibitem{RuppeinerIII} P. Salamon, J. Nulton, and E. Ihrig, J. Chem. Phys. 
\textbf{80}, 436 (1984).

\bibitem{QuevedoI} H. Quevedo, Gen. Relativ. Gravit. \textbf{40}, 971 (2008).

\bibitem{QuevedoII} H. Quevedo, and A. Sanchez, JHEP \textbf{09}, 034 (2008).

\bibitem{HPEM} S. H. Hendi, S. Panahiyan, B. Eslam Panah, and M. Momennia,
Eur. Phys. J. C \textbf{75}, 507 (2015).

\bibitem{HPEM1} S. H. Hendi, S. Panahiyan, and B. Eslam Panah, Adv. High
Energy Phys. \textbf{2015}, 743086 (2015).

\bibitem{HPEM2} S. H. Hendi, A. Sheykhi, S. Panahiyan, and B. Eslam Panah,
Phys. Rev. D \textbf{92}, 064028 (2015).

\bibitem{HPEM3} S. H. Hendi, B. Eslam Panah, and S. Panahiyan, JHEP \textbf{%
05}, 029 (2016).

\bibitem{GTD1} S. A. H. Mansoori, and B. Mirza, Eur. Phys. J. C \textbf{74},
2681 (2014).

\bibitem{GTD2} S. A. H. Mansoori, B. Mirza, and M. Fazel, JHEP \textbf{04},
115 (2015).

\bibitem{GTD3} R. Banerjee, and B. R. Majhi, S. Samanta, Phys. Lett. B 
\textbf{767}, 25 (2017).

\bibitem{GTD4} K. Bhattacharya, and B. R. Majhi, Phys. Rev. D \textbf{95},
104024 (2017).

\bibitem{ModMaxII} B. P. Kosyakov, Phys. Lett. B \textbf{810}, 135840 (2020).

\bibitem{Rcont1} A. de la Cruz-Dombriz, A. Dobado, and A. L. Maroto, Phys.
Rev. D \textbf{80}, 124011 (2009).

\bibitem{Rcont2} T. Moon, Y. S. Myung, and E. J. Son, Gen. Relativ. Gravit. 
\textbf{43}, 3079 (2011).

\bibitem{Cognola2005} G. Cognola, E. Elizalde, S. Nojiri, S. D. Odintsov,
and S. Zerbini, JCAP \textbf{02}, 010 (2005).

\bibitem{AMDI} A. Ashtekar, and A. Magnon, Class. Quantum Grav. \textbf{1},
L39 (1984).

\bibitem{AMDII} A. Ashtekar, and S. Das, Class. Quantum Grav. \textbf{17},
L17 (2000).

\bibitem{Rodrigues2012} M. E. Rodrigues, and Z. A. A. Oporto, Phys. Rev. D 
\textbf{85}, 104022 (2012).

\bibitem{Han2012} Y. W. Han, and G. Chen, Phys. Lett. B \textbf{714}, 127
(2012).

\bibitem{Azreg2014} M. Azreg-Ainou, Eur. Phys. J. C \textbf{74}, 2930 (2014).

\bibitem{Suresh2015} J. Suresh, R. Tharanath, and V. C. Kuriakose, JHEP 
\textbf{01}, 019 (2015).

\bibitem{Mo2016} J. -X. Mo, G. -Q. Li, and Y. -C. Wu, JCAP \textbf{04}, 045 
(2016).

\bibitem{Quevedo2016} H. Quevedo, M. N. Quevedo, and A. Sanchez, Phys. Rev.
D \textbf{94}, 024057 (2016).

\bibitem{Channuie2018} P. Channuie, and D. Momeni, Phys. Lett. B \textbf{785}%
, 309 (2018).

\bibitem{Bhattacharya2019} K. Bhattacharya, S. Dey, B. R. Majhi, and S.
Samanta, Phys. Rev. D \textbf{99}, 124047 (2019).

\bibitem{HPEM4} B. Eslam Panah, Phys. Lett. B \textbf{787}, 45 (2018).

\bibitem{Mansoori2019} S. A. H. Mansoori, and B. Mirza, Phys. Lett. B 
\textbf{799}, 135040 (2019).

\bibitem{Upadhyay2021} S. Upadhyay, S. Soroushfar, and R. Saffari, Mod.
Phys. Lett. A \textbf{36}, 2150212 (2021).

\bibitem{HPEM5} Kh. Jafarzade, J. Sadeghi, B. Eslam Panah, and S. H. Hendi,
Ann. Phys. \textbf{432}, 168577 (2021).

\bibitem{Mahmoudi2023} S. Mahmoudi, Kh. Jafarzade, and S. H. Hendi, Turk. J.
Phys. \textbf{47}, 214 (2023).

\bibitem{Makarenko2014} A. N. Makarenko, S. D. Odintsov, and G. J. Olmo, Phys.
Rev. D \textbf{90}, 024066 (2014).

\bibitem{Makarenkoa2014} A. N. Makarenko, S. D. Odintsov, and G. J. Olmo,
Phys. Lett. B \textbf{734}, 36 (2014).
\end{thebibliography}
\end{document}